%% file: mainFinal.tex
\documentclass[a4paper,twocolumn,11pt,accepted=2025-12-08]{quantumarticle}
\pdfoutput=1
\usepackage[utf8]{inputenc}
\usepackage[english]{babel}
\usepackage[T1]{fontenc}
\usepackage{amsmath}
\usepackage{hyperref}
\usepackage{tikz}
\usepackage{lipsum}

\newcommand\hmmax{0}
\newcommand\bmmax{0}

\usepackage{graphicx}
\usepackage{tikz}
\usepackage{pgfplots}
\usepackage{xcolor}
\usepackage{color, colortbl}
\usepackage{amsmath}
\usepackage{bbm}
\usepackage{amsfonts, amssymb}
\usepackage{mathtools}
\usepackage{cite}
\usepackage[nolist]{acronym}
\usepackage{amsmath}
\DeclareMathOperator{\tr}{tr}

\usepackage{booktabs} 		
\usepackage{diagbox}		
\usepackage{upgreek}
\usepackage{bbm}
\usepackage{Files/bm}
\usepackage{Files/winsnotation}
\usepackage{Files/Symbols_OPL}
\usepackage{Files/LVcolours} 

\usepackage[ruled,vlined,noresetcount]{algorithm2e}
\usepackage{setspace}	 	
\usepackage{comment}
\usepackage{physics}

\usepackage{array}
\newcolumntype{C}[1]{>{\centering\arraybackslash}p{#1}}

\makeatletter
\def\endthebibliography{%
  \def\@noitemerr{\@latex@warning{Empty `thebibliography' environment}}%
  \endlist
}
\makeatother
\usepackage{amsthm}
\theoremstyle{definition}
\newtheorem{definition}{Definition}

\newcommand{\px}{p_{\mathrm{X}}}
\newcommand{\py}{p_{\mathrm{Y}}}
\newcommand{\pz}{p_{\mathrm{Z}}}

\newcommand{\PauliX}{\M{X}}
\newcommand{\PauliY}{\M{Y}}
\newcommand{\PauliZ}{\M{Z}}

\newcommand{\PauliXped}[1]{\M{X}_{{#1}}}
\newcommand{\PauliYped}[1]{\M{Y}_{{#1}}}

\newcommand{\eg}{e_{\mathrm g}}
\newcommand{\eZ}{e_{\mathrm Z}}
\newcommand{\Syndrome}[0]{\V{s}}

\pgfplotsset{compat=1.17}

\begin{document}

\title{Performance Analysis of Quantum CSS Error-Correcting  Codes via MacWilliams Identities}

\author{Diego Forlivesi}
\email{diego.forlivesi2@unibo.it}
\author{Lorenzo Valentini}
\email{lorenzo.valentini13@unibo.it}
\author{Marco Chiani}
\email{marco.chiani@unibo.it}
\affiliation{Department of Electrical, Electronic, and Information Engineering ``Guglielmo Marconi'' and CNIT/WiLab, University of Bologna, 40136 Bologna, Italy.}
\maketitle

\input{Files/Acronimi_SICMMA.tex}
\setcounter{page}{1}

\begin{abstract}
We analyze the performance of quantum stabilizer codes, one of the most important classes for practical implementations, on both symmetric and asymmetric quantum channels. To this aim, we first derive the weight enumerator (WE) for the undetectable errors based on the quantum MacWilliams identities. The WE is then used to evaluate tight upper bounds on the error rate of CSS quantum codes with \acl{MW} decoding.
For surface codes we also derive a simple closed form expression of the bounds over the depolarizing channel.
We introduce a novel approach that combines the knowledge of WE with a logical operator analysis, allowing the derivation of the exact asymptotic error rate for short codes. 
For example, on a depolarizing channel with physical error rate $\rho \to 0$, the logical error rate $\rho_\mathrm{L}$ is asymptotically $\rho_\mathrm{L} \approx 16 \rho^2$ for the $[[9,1,3]]$ Shor code,  $\rho_\mathrm{L} \approx 16.3 \rho^2$ for the $[[7,1,3]]$ Steane code, $\rho_\mathrm{L} \approx 18.7 \rho^2$ for the $[[13,1,3]]$ surface code, and $\rho_\mathrm{L} \approx 149.3 \rho^3$ for the $[[41,1,5]]$ surface code. For larger codes our bound provides $\rho_\mathrm{L} \approx 1215 \rho^4$ and $\rho_\mathrm{L} \approx 663 \rho^5$ for the $[[85,1,7]]$ and the $[[181,1,10]]$ surface codes, respectively.
Finally, we extend our analysis to include realistic, noisy syndrome extraction circuits by modeling error propagation throughout gadgets. This enables estimation of logical error rates under faulty measurements.
The performance analysis serves as a design tool for developing fault-tolerant quantum systems by guiding the selection of quantum codes based on their error correction capability. 
Additionally, it offers a novel perspective on quantum degeneracy, showing it represents the fraction of non-correctable error patterns shared by multiple logical operators.
\end{abstract}

\section{Introduction}

The exploitation of the unique features of quantum mechanics has opened new perspectives on how we can sense, process, and communicate information \cite{Kim:08,Pre:18,NdoPetGri:19,NAP:19,WehElkHan:18,CacCalVan:20}. From an engineering point of view, there are many challenges to solve, calling for both theoretical and experimental research studies. The aim is to progress towards the already known possible applications of quantum information technologies, as well as those currently still unforeseen, that will arise when  practical implementations become available. 
One of the main challenges is how to deal with the noise caused by unwanted interaction of the quantum information with the environment \cite{Sho:95,Laf:96,Got:96,Kni:97,FleShoWin:08,NieChu:10,RehShi:21}. Quantum error correcting codes, where a redundant representation of quantum states protects from certain types of errors, are therefore of paramount importance for quantum computation, quantum memories, and quantum communication systems \cite{Ste:96a,Got:09,Ter:15,MurLiKim:16,Rof:19,Bab:19,ChiConWin:20,ZorDePGio:23, BluDolEve:23, Cam:17, YodKim:17, BriBro:09}. 

Quantum error correction is made difficult by the laws of quantum mechanics which imply that qubits cannot be copied or measured without perturbing state superposition. 
Moreover, there is a continuum of errors that could occur on a qubit. 
However, it has been shown that to correct an arbitrary qubit error, assuming it remains within the computational space (i.e., neglecting leakage), it is sufficient to consider error correction over the discrete set of Pauli operators
\cite{Kni:97,Got:09,NieChu:10}.
Indeed, assume that a generic continuous phase rotation has occurred on a qubit. 
Applying the standard syndrome-extraction circuit, we first interact with ancilla qubits to determine the error syndrome associated with these errors. 
Next, we measure the ancilla qubits. 
This measurement collapses the superposition, yielding one of two results: either no error has occurred, or a Pauli error has been identified \cite{Got:96, Got:09}. 
Thus, we can consider in general a Pauli channel, commonly used in literature for analyzing stabilzer-based \ac{QECC} \cite{Kni:97, FowMarMar:12 ,Got:09,NieChu:10}. 
This channel introduces qubit Pauli errors $\PauliX$, $\PauliY$, and $\PauliZ$ with probabilities $\px$, $\py$, and $\pz$, respectively, and leaving the qubit intact with probability $1-\rho$, where $\rho = \px + \py + \pz$. 
A special case of this model is the so-called depolarizing (or symmetric) channel for which $\px = \py = \pz = \rho/3$. Quantum error-correcting codes for this channel are thus designed to protect against equiprobable Pauli errors 
\cite{Sho:95,Laf:96,Ste:96a,NieChu:10,OstOrsLaz:22}. 
However, depending on the technology adopted for the system implementation, the different types of Pauli error can have different probabilities of occurrence, leading to polarizing (or asymmetric) quantum channels \cite{IofMez:07,SarKlaRot:09,LayCheCap:20,ChiVal:20a}. 
Let us now consider channels that arise in the study of decoherence in concrete physical systems, specifically the combined \emph{amplitude damping and dephasing channels}.
Key parameters for the noise process underlying this channel are the relaxation time $T_1$ and dephasing time $T_2$. Using a technique called twirling, often employed in quantum information theory, it is possible to transform this channel into the corresponding Pauli channel by conjugating the channel with Pauli matrices and averaging the results \cite{SarKlaRot:09, DiV:02, Eme:05, Dan:09}. 
The significance of the twirling method lies in the fact that a correctable code for the twirled channel will also be a correctable code for the original channel \cite{Iol:24, Sil:08, Mar:20}.
For instance, given a combined amplitude damping and dephasing channel, the associated Pauli-twirled channel has $\px = \pz = (1 - e^{-\frac{t}{T_1}}) / 4$ and $\pz = (1 - 2\px - e^{-\frac{t}{T_2}} )/ 2$.
In this context, the proposed analysis can be easily extended to more realistic channel models.

In this paper we provide an analytical evaluation of the performance of generic stabilizer codes, including surface codes \cite{Got:96,NieChu:10,BraKit:98,FowMarMar:12,AchRajAle:22, KriSebLac:22, ZhaYouYe:22, Gid:21, Aar:04}. 
These codes can be interpreted as structured realizations of \ac{QLDPC} codes, characterized by low-weight stabilizer generators and a planar architectural layout.
Stabilizer codes are indeed the most important class of \acp{QECC} for practical implementations. 
Despite their importance, the theoretical performance of these codes has been investigated in the literature only partially, and mainly in terms of accuracy threshold over symmetric channels for fault-tolerant quantum computing \cite{Got:96,Pre:98,NieChu:10,BraKit:98,FowMarMar:12,BraSerSuc:14,AtaTucBar:21}. 
We propose a framework for the performance investigation of stabilizer codes by means of the quantum MacWilliams identities.
In \cite{Ash00:MacwPartI, Ash00:MacwPartII} these identities have been employed to derive theoretical bounds for quantum error detection. 
Here, we exploit the undetectable errors weight enumerator to provide upper bounds on the logical error rate for \ac{CSS} codes \cite{CalSho:96,Ste:96}.
Moreover, we develop a logical operator analysis leading to exact expressions for the logical error rates, assuming complete decoders (decoders that always attempt to correct the error).
Specifically, we analyze \acf{MW} decoding, which finds the lowest weight error consistent with the syndrome \cite{ Pou:06, Hsi:11, IyePavPou:15}. 
This choice is driven by practical considerations related to the complexity of the decoder~\cite{HigGid:23, Mu25:Relay}.
In particular, for surface codes, an instance of \ac{QLDPC} codes, the decoder is typically realized with the help of the blossom algorithm for finding a \ac{MWPM} in a graph \cite{Edm:65, DenKitLan:02, Hig:21, SkoBro:23}. 
Although \ac{ML} decoders exist, such as the \ac{MPS} decoder \cite{BraSerSuc:14}, their higher latency introduces decoherence effects that outweigh the benefits gained from improved decoding performance.
Since \ac{MW} decoding is suboptimal with respect to \ac{ML} decoding, our upper bounds remain applicable; however, their tightness may be limited for codes with high degeneracy.
For non-\ac{QLDPC} codes, where degeneracy could be difficult to assess, the tightness of the bounds cannot be guaranteed.
Currently, long non-\ac{QLDPC} codes are of little practical use owing to their implementation complexity, which further justifies the focus of our analysis.

The performance analysis presented in this work is conducted for both symmetric and asymmetric models of quantum channel errors.
In practical quantum systems, however, syndrome extraction is a critical yet error-prone component of quantum error correction. 
Measurements are inherently noisy and typically require repetition to ensure reliability. 
Additionally, faults during extraction can propagate, causing high-weight correlated errors \cite{Got:09}.
To address these challenges, we introduce a framework that models the full syndrome extraction process, incorporating gate-specific noise and measurement imperfections.
This enables the estimation of logical error rates under realistic circuit-level noise assumptions.
Our analysis provides a practical valuable tool for the development of fault-tolerant quantum systems, helping guide the selection of specific quantum codes based on their error correction capability. 

Furthermore, we provide advancements in the understanding of quantum degeneracy. 
Quantum degeneracy typically refers to the non-uniqueness of the error-correcting syndrome, meaning that different errors can produce the same syndrome. 
Since degeneracy is related to physical errors and has no direct counterpart in classical information theory, it is often challenging to study, especially in the construction of quantum codes.
Here, we offer a completely new perspective by showing that the degeneracy of a quantum code represents the fraction of error patterns with weight greater than $t = \lfloor(d-1)/2 \rfloor$ that are shared by more than one logical operator. This characteristic depends on the structure of the code and can therefore serve as a useful guideline in the design of new quantum codes.

The key contributions of the paper can be summarized as follows:
\begin{itemize}
    \item we derive the weight enumerator for the undetectable errors $L(z)$ of arbitrary stabilizer codes via MacWilliams identities;
    \item we derive theoretical upper bounds for the error correction capability of \ac{CSS} stabilizer codes;
    \item we derive closed form expressions for the $L(z)$ coefficients which significantly impact the performance of surface codes for any code distance; 
    \item we derive the exact performance of stabilizer codes under \ac{MW} decoding, including surface codes under \ac{MWPM} decoding, over symmetric and asymmetric channels;
    \item we extend the analysis to include realistic, noisy syndrome extraction circuits, providing a method for estimating logical error rates under circuit-level noise models;
    \item we introduce a novel perspective on quantum degeneracy, analyzing its influence on the error correction capability of a quantum code.
\end{itemize}

As relevant examples, we give the expressions for the performance of the Shor code, the Steane code, and of surface codes of arbitrary size. 

This paper is organized as follows. Section~\ref{sec:preliminary} introduces preliminary concepts and models together with some background material. Sections~\ref{sec:TheoPerfBoundedDist} and~\ref{sec:TheoPerfComplete} provide the analytical investigation of \ac{QECC} with bounded distance, and with complete decoding. In Section~\ref{sec:MacWilliams_intro} we derive the \ac{WE} for the undetectable errors from MacWilliams identities and apply it to the evaluation of the logical error rate of arbitrary stabilizer codes. 
In Section~\ref{Sec_syndromeExtr}, we analyze the impact of noisy syndrome measurement extraction circuits.
Numerical results are discussed in Section~\ref{sec:NumRes}. 
\section{Preliminaries and Background}
\label{sec:preliminary}

\subsection{Quantum Stabilizer Codes}
\label{subsec:QEC}
A qubit is an element of the two-dimensional Hilbert space $\mathcal{H}^{2}$, with basis $\ket{0}$ and $\ket{1}$ \cite{NieChu:10}. 
The Pauli operators $\M{I}, \M{X}, \M{Z}$, and $\M{Y}$, are defined by  $\M{I}\ket{a}=\ket{a}$, $\M{X}\ket{a}=\ket{a\oplus 1}$, $\M{Z}\ket{a}=(-1)^a\ket{a}$, and $\M{Y}\ket{a}=i(-1)^a\ket{a\oplus 1}$ for $a \in \lbrace0,1\rbrace$. These operators either commute (e.g. $\M{I}\M{X}=\M{X}\M{I}$) or anticommute (e.g. $\M{X}\M{Z}=-\M{Z}\M{X}$) with each other. Also, apart from an overall factor $\pm 1, \pm i$, the composition of two Pauli produces another Pauli (e.g. $\M{X}\M{Y} = i\M{Z}$). Thus, all the Pauli operators, together with multiplicative factors $\pm 1, \pm i$ constitute a group, indicated as $\mathcal{G}_1$. Similarly, all Pauli operators on $n$ qubits together with multiplicative factors $\pm 1, \pm i$ form the $\mathcal{G}_n$ Pauli group \cite{Got:09,NieChu:10}. 
We indicate with $[[n,k,d]]$ a \ac{QECC} with minimum distance $d$, that encodes $k$ information qubits  $\ket{\varphi}$ (called logical qubits) into a codeword of $n$ qubits  $\ket{\psi}$ (called data or physical qubits), allowing the decoder to correct all patterns up to $t = \lfloor(d-1)/2 \rfloor$ errors (and some patterns of more errors). To simplify our analysis, we consistently adopt the assumption that $d$ is an odd number. The codewords will be assumed equiprobable in the following. 
Using the stabilizer formalism, we start by choosing $n-k$ independent and commuting operators $\M{G}_i \in \mathcal{G}_n$, called stabilizer generators (or simply generators).   
The subgroup of $\mathcal{G}_n$ generated by all combinations of the $\M{G}_i \in \mathcal{G}_n$ is a stabilizer, indicated as $\mathcal{S}$. 
The code $\mathcal{C}$ is the set of quantum states $\ket{\psi}$ stabilized by $\mathcal{S}$, i.e., satisfying 
$\M{S}\ket{\psi}=\ket{\psi} \, \forall \M{S} \in \mathcal{S}$, or, equivalently, 
$\M{G}_i \ket{\psi}=\ket{\psi},\, i=1, 2, \ldots, n-k$. 
 For a subgroup $\mathcal{H}$ of a group $\mathcal{G}$ we indicate with
$\mathcal{N}(\mathcal{H})$ the normalizer, and with $\mathcal{C}(\mathcal{H})$ the centralizer. 
The centralizer and the normalizer of the stabilizer $\mathcal{S}$ are coincident, $\mathcal{N}(\mathcal{S})=C(\mathcal{S})$. 

Assume a codeword $\ket{\psi} \in \mathcal{C}$ is affected by a channel error. 
Measuring the received state according to the generators $\M{G}_i$ with the aid of ancilla qubits, the error collapses on a discrete set of possibilities represented by the Pauli operators $\M{E}\in \mathcal{G}_n$ \cite{Got:09}. We call this $\M{E}$ a Pauli error. The weight of an error $\M{E} \in \mathcal{G}_n$ is the number of single qubits Pauli operators which are not equal to the identity. 
For example, the error $\M{E} = \PauliXped{2} \PauliYped{3}$ has weight two, with $\M{X}$ occurred on the second qubit and $\M{Y}$ occurred on the third qubit (we implicitly mean that the others qubits see the Pauli identity operator). 
The measurement procedure over the ancilla qubits results in a quantum error syndrome $\Syndrome(\M{E})=(s_1, s_2, \ldots,s_{n-k})$, with each $s_i =0$ or $1$ depending on $\M{E}$ commuting or anticommuting with $\M{G}_i$, respectively \cite{Got:09}. In the following, we will refer to ancillas measuring $s_i = 1$ as \emph{defects}.
Note that an error $\M{E}\in \mathcal{S}$ has no effect on a codeword since in this case $\M{E} \ket{\psi}=\ket{\psi}$. A minimum weight decoder will infer the most probable error $\M{\hat{E}}\in \mathcal{G}_n$ compatible with the measured syndrome.  
Within the category of stabilizer codes, among the earliest introduced quantum codes were the $[[9,1,3]]$ Shor code \cite{Sho:95}, with generators 
 \begin{equation*}
\arraycolsep=10pt
\begin{array}{lllllll} 
\M{G}_1 = \mathbf{Z}_1 \mathbf{Z}_2 & 
\M{G}_2 = \mathbf{Z}_2 \mathbf{Z}_3 & \\
\M{G}_3 = \mathbf{Z}_4 \mathbf{Z}_5 & 
\M{G}_4 = \mathbf{Z}_5 \mathbf{Z}_6 & \\
\M{G}_5 = \mathbf{Z}_7 \mathbf{Z}_8 & 
\M{G}_6 = \mathbf{Z}_8 \mathbf{Z}_9 & \\
\M{G}_7 = \mathbf{
X}_1 \mathbf{X}_2 \mathbf{X}_3 \mathbf{X}_4 \mathbf{X}_5 \mathbf{X}_6  & \\ 
\M{G}_8 = \mathbf{
X}_4 \mathbf{X}_5 \mathbf{X}_6 \mathbf{X}_7 \mathbf{X}_8 \mathbf{X}_9 ,  &
\end{array} 
\end{equation*} 
the $[[7,1,3]]$ Steane code \cite{SteAnd:96}, with generators 
\begin{equation*}
\arraycolsep=2.8pt
\begin{array}{lllllll} 
\M{G}_1 = \mathbf{X}_1 \mathbf{X}_3 \mathbf{X}_5 \mathbf{X}_7 &
\M{G}_2 = \mathbf{X}_2 \mathbf{
X}_3 \mathbf{X}_6 \mathbf{X}_7 & \\
\M{G}_3 = \mathbf{X}_4 \mathbf{
X}_5 \mathbf{X}_6 \mathbf{X}_7 & 
\M{G}_4 = \mathbf{Z}_1 \mathbf{Z}_3 \mathbf{Z}_5 \mathbf{Z}_7 & \\
\M{G}_5 = \mathbf{Z}_2 \mathbf{
Z}_3 \mathbf{Z}_6 \mathbf{Z}_7 & 
\M{G}_6 = \mathbf{Z}_4 \mathbf{
Z}_5 \mathbf{Z}_6 \mathbf{Z}_7 , & \\
\end{array} 
\end{equation*} 
and the $[[5,1,3]]$ perfect code \cite{LafRayMiq:96} with generators
\begin{equation*}
\arraycolsep=2.8pt
\begin{array}{lllllll} 
\M{G}_1 = \mathbf{X}_1 \mathbf{Z}_2 \mathbf{Z}_3 \mathbf{X}_4 & \quad
\M{G}_2 = \mathbf{X}_2 \mathbf{
Z}_3 \mathbf{Z}_4 \mathbf{X}_5 & \\
\M{G}_3 = \mathbf{
X}_1 \mathbf{X}_3 \mathbf{Z}_4 \mathbf{Z}_5 & \quad 
\M{G}_4 = \mathbf{Z}_1 \mathbf{X}_2 \mathbf{X}_4 \mathbf{Z}_5. &
\end{array} 
\end{equation*} 

A possible channel model is one characterized by errors occurring independently and with the same statistic on the individual qubits of each codeword. In this model, the error on each physical qubit can be $\M{X}$, $\M{Z}$ or $\M{Y}$ with probabilities $\px$, $\pz$, and $\py$, respectively. The probability of a generic error on a physical qubit is $\rho = \px + \pz + \py$.
Two important models are the \emph{depolarizing channel} where $\px = \pz = \py = \rho / 3$, and the \emph{phase flip channel} where $\rho=\pz$, $\px = \py=0$. 
We will also consider more general asymmetric channels with the constraint $\px=\py$, therefore completely characterized by the bias parameter $A = 2\pz /(\rho - \pz)$. Note that for $A=1$ we have the depolarizing channel, and for $A \to \infty$ we have the phase flip channel.

In the following, we will also adopt the notation $[[n, k,d_\mathrm{X}/d_\mathrm{Z}]]$ for asymmetric codes able to correct all patterns up to $t_\mathrm{X} = \lfloor(d_\mathrm{X}-1)/2\rfloor$ Pauli $\M{X}$ errors and $t_\mathrm{Z} = \lfloor(d_\mathrm{Z}-1)/2\rfloor$ Pauli $\M{Z}$ errors.

\subsection{Quantum Topological Codes}

\begin{figure}[t]
	\centering
    \includegraphics[width = \columnwidth]{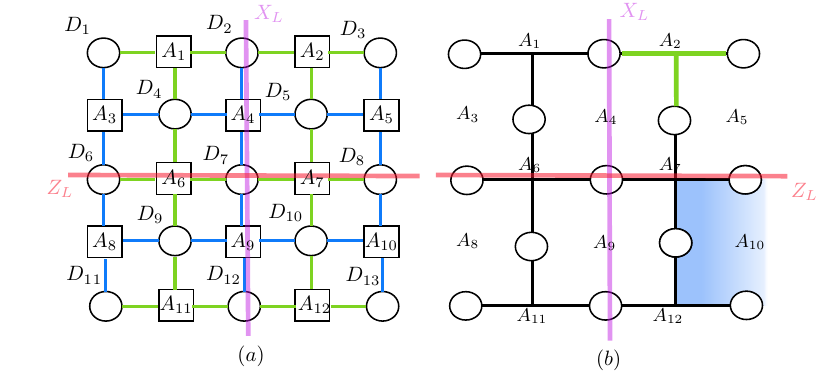}
	\caption{$[[ 13,1,3 ]]$ Surface code. (a) Actual lattice with data qubits $D$ (circles) and ancillae $A$ (squares). (b) Simplified representation with $\M{X}$ generators corresponding to sites and $\M{Z}$ generators to plaquettes. A smooth edge ($A_2$) and a rough edge ($A_{10}$) are depicted in green and blue, respectively. Examples of logical operators are drawn on the lattice. }
	\label{Fig:surface}
\end{figure}

One of the most important families of stabilizer \ac{QECC} is that of  \emph{topological} codes. The general design principle behind these codes is that they are built by patching together repeated elements. Using this kind of approach, they can be easily scaled in size in order to increase the distance of the code, still guaranteeing commutativity of the generators. 
With regard to the actual implementation, these codes have a great intrinsic advantage. 
In fact, they require only nearest-neighbor interactions\cite{Rof:19}. 
The most important codes within this category are the \emph{surface} codes, in which all the check operators are local and the qubits are arranged on planar sheets. 
Specifically, the following will focus on unrotated surface codes.
Each stabilizer is associated either with one of the sites or one of the cells that are called \lq\lq plaquettes\rq\rq \cite{HorFowDev:12}. 
The stabilizer's generators in the interior are four-qubits plaquette or site operators, while the ones at the boundaries are three-qubits operators.  Along a plaquette or \lq\lq rough\rq \rq \; edge, each generator is  a three-qubits operator $\M{Z}^{\otimes 3}$, while, along a site edge or \lq\lq smooth\rq\rq \; edge, each generator is a three-qubits operator $\M{X}^{\otimes 3}$. 
The entire lattice is able to encode one logical qubit ($k=1$). 
It can be shown that in a code with distance $d$, the lattice has $d^2 + (d-1)^2$ physical qubits \cite{DenKitLan:02}. For example, two equivalent graphical representations of the resulting lattice for the $[[13,1,3]]$ surface code are shown in Fig.~\ref{Fig:surface}. For this surface code, the  generators are
\begin{equation*}
\arraycolsep=2.5pt
\begin{array}{lllllll} 
\M{G}_1 = \mathbf{X}_1 \mathbf{X}_2 \mathbf{X}_4 &
\M{G}_2 = \mathbf{X}_2 \mathbf{X}_3 \mathbf{X}_5 & \\
\M{G}_3 = \mathbf{Z}_1 \mathbf{Z}_4 \mathbf{Z}_6 &
\M{G}_4 = \mathbf{Z}_2 \mathbf{Z}_4 \mathbf{Z}_5 \mathbf{Z}_7 & \\ 
\M{G}_5 = \mathbf{Z}_3 \mathbf{Z}_5 \mathbf{Z}_8 \\
\M{G}_6 = \mathbf{X}_4 \mathbf{X}_6 \mathbf{X}_7 \mathbf{X}_9 &
\M{G}_7 =  \mathbf{X}_5 \mathbf{X}_7 \mathbf{X}_8 \mathbf{X}_{10} & \\
\M{G}_8 = \mathbf{Z}_6 \mathbf{Z}_9 \mathbf{Z}_{11} &
\M{G}_9 = \mathbf{Z}_7 \mathbf{Z}_9 \mathbf{Z}_{10} \mathbf{Z}_{12} & \\ 
\M{G}_{10} = \mathbf{Z}_8 \mathbf{Z}_{10} \mathbf{Z}_{13} \\
\M{G}_{11} = \mathbf{X}_9 \mathbf{X}_{11} \mathbf{X}_{12} &
\M{G}_{12} = \mathbf{X}_{10} \mathbf{X}_{12} \mathbf{X}_{13} \,. &
\end{array} 
\end{equation*} 
The logical logical $\M{Z}_L$ operator can be chosen as a tensor product of $\M{Z}$'s acting on a chain of qubits running from a rough edge to the one at the opposite side of the lattice. Similarly, the $\M{X}_L$ logical operator will be the tensor product of $\M{X}$'s acting on a chain running from a smooth link to the correspondent one at the other side of the dual lattice. 
Another important feature is that a simple complete decoder exists for surface codes: the \ac{MWPM} \cite{Hig:21}. This decoder connects pairs of defects in the shortest way. 
Moreover, any set of $\M{Z}$ ($\M{X}$) operators which form a closed loop on the edges of the lattice is contained in the stabilizer. Hence, if the correction operator applied by the decoder together with the channel error closes a loop, the error is correctly recovered. As a consequence, surface codes can also easily correct a large number of errors with a weight larger than $t = \lfloor(d-1)/2 \rfloor$. 

In general, a decoding error occurs every time that the concatenation of actual channel error and correction operator realizes a logical operator, so the whole chain operator crosses the lattice from boundary to boundary, realizing a logical operator. For instance, consider a two qubits error $\M{Z}_2\M{Z}_3$ for the code in Fig.~\ref{Fig:surface}. In this scenario, the ancilla qubit $A_1$ is the only one that changes its state, hence the decoder assumes (wrongly) an error $\M{Z}$ on data qubit $D_1$. It can be easily noticed that the whole operator applied to the lattice represents the logical $\M{Z}_L$.

Rectangular surface codes, with differing horizontal and vertical dimensions, can be effectively employed in asymmetric channels. For instance, in hardware implementations, $\M{Z}$ errors tend to occur more frequently \cite{IofMez:07,ChiVal:20a,AzaLipMah:22}. In such scenarios, the $[[23,1,3/5]]$ surface code is a viable option, where the horizontal direction is two qubits longer than the vertical one. This configuration results in the logical $\M{Z}_L$ operator being a chain of 5 qubits, in contrast to the $[[13,1,3]]$ code, where it spans only 3 qubits. Despite requiring a lattice of 23 qubits, this code has a distance of $d_{Z} = 5$, enabling it to correct weight two $\M{Z}$ errors.

\subsection{Quantum Codes with Bounded Distance Decoding}
\label{sec:TheoPerfBoundedDist}
The codeword error probability, $\rho_\mathrm{L}$, also called logical error probability, is defined as the probability that the decoder does not correct the errors introduced by the quantum channel.

Let us first assume an $[[n,k,d]]$ \ac{QECC} together with a decoder which corrects up to $t=\lfloor (d-1) /2 \rfloor $ generic errors (i.e., $\PauliX$, $\PauliZ$, or $\PauliY$) per codeword, and no others.  For this \ac{BD} decoder, the logical error probability is simply
\begin{equation}
\label{eq:PeGen}
\rho_\mathrm{L} = 1 - \sum_{j = 0}^{t} \binom{n}{j} \rho^j \, (1-\rho)^{n-j}
\end{equation}
that, for $\rho \ll 1$, can be approximated as
\begin{align}
\label{eq:error_probBDasy}
\rho_\mathrm{L} 
&\simeq \binom{n}{t+1}\rho^{t+1} \,.
\end{align}
We can see that the slope in the log-log plot of the logical error probability $\rho_\mathrm{L}$ vs.~the physical error probability $\rho$ is $t+1$.

The error probability analysis has been recently generalized to asymmetric \acp{QECC} assuming a decoder able to correct up to $\eg$ generic errors plus up to $\eZ$ Pauli $\PauliZ$ errors per codeword, and no others. %
In this case, weighting each pattern with the corresponding probability of occurrence, the bounded distance decoding error rate \eqref{eq:PeGen} over an asymmetric quantum channel with arbitrary  $\px$, $\py$, and $\pz$ becomes \cite{ChiVal:20a}
\begin{align}
\label{eq:PeGen2}
 \nonumber \rho_\mathrm{L} 
= \,\, &1 - \sum_{j = 0}^{\eg+\eZ} \binom{n}{j}(1-\rho)^{n-j} \\
&\times \sum_{i = (j-\eg)^+}^{j}\binom{j}{i} \, \pz^i \, \left(\rho-\pz\right)^{j-i}
\end{align}
where $(x)^+=\max\{x,0\}$. For a channel with asymmetry parameter $A = 2\pz /(\rho - \pz)$, noting that $(j-\eg)^+ = 0$ if $j \le e_\mathrm{g}$ and using the binomial theorem, the expression in \eqref{eq:PeGen2} can be rewritten 
\begin{align}
\label{eq:PeAsym}
\rho_\mathrm{L} =&\: 1 - \sum_{j = 0}^{\eg+\eZ} \alpha_j \binom{n}{j}\rho^j  (1-\rho)^{n-j}
\end{align}
where
\begin{align}
\label{eq:PeGenxi_2}
\alpha_j = 
&\begin{dcases}
1 & \text{if}~j\le\eg\\ 
\left(\frac{2}{A+2} \right)^j \sum_{i = j-\eg}^{j}\binom{j}{i} \left( \frac{A}{2} \right) ^i & \text{otherwise} . \,\\
\end{dcases} 
\end{align} 
Note that, when the code is symmetric $\eZ=0$, and \eqref{eq:PeAsym} reduces to  \eqref{eq:PeGen}. 

Similarly to the approximation done in \eqref{eq:error_probBDasy} derived from \eqref{eq:PeGen} when $\rho \ll 1$, considering the most significant terms in \eqref{eq:PeAsym}, the asymptotic slope analysis can be extended to asymmetric codes with $e_\mathrm{Z} \geq 1$ as
\begin{align}
\label{eq:PeAsymApprox}
\rho_\mathrm{L} &\simeq 
\begin{dcases}
\binom{n}{e_\mathrm{g}+1}\,\left( \frac{2\rho}{A+2}\right)^{e_\mathrm{g}+1} & 1 \le A < \infty\\
\binom{n}{e_\mathrm{g} + e_\mathrm{Z} +1}\,\rho^{e_\mathrm{g} + e_\mathrm{Z}+1} & A \to \infty\,.\\
\end{dcases}
\end{align}
We observe that in this case, the asymptotic slope is $e_\mathrm{g} + 1$ for all finite $A$. On the other hand, considering a phase flip channel ($A \to \infty$) the slope becomes $e_\mathrm{g} + e_\mathrm{Z} + 1$. 

\section{Quantum Codes with Minimum Weight Decoding}
\label{sec:TheoPerfComplete}

Let us now assume to have a decoder that always outputs a codeword (complete decoder). 
This could allow correcting also some (but not all) error patterns which are not correctable with bounded distance decoding. Specifically, we analyse the \ac{MW} decoder, which can be implemented as the \ac{MWPM} decoder for surface codes. 

\begin{definition}
We indicate with $\beta_j$ the fraction of errors of weight $j$ that can be corrected by a complete decoder. 
\end{definition}
Note that $\beta_j$ depends in general on the code structure, on the decoder, and on the channel asymmetry parameter $A$.  

\begin{definition}[Error class]
    We state that two error patterns are in the same class if they have the same Pauli weight with respect to each Pauli operator, i.e., they contain the same number of $\M{X}$, $\M{Y}$, and $\M{Z}$ errors, respectively.
\end{definition}

In general, the logical error probability of an error-correcting code of length $n$ is
\begin{align}
    \rho_L = \sum_{\M{E} \in \mathcal{C}_n} \Prob{\text{error}|\M{E}}\Prob{\M{E}}
\end{align}
where $\mathcal{C}_n$ is the set of all possible error classes over $n$ qubits, $\Prob{\text{error}|\M{E}}$ is the probability to have an error given the particular error class $\M{E}$, and $\Prob{\M{E}}$ is the occurrence probability of $\M{E}$.
Since the probability $\Prob{\text{error}|\M{E}}$ depends only on how many $\M{X}$, $\M{Y}$, and $\M{Z}$ the error $\M{E}$ contains, we define $f_j(i,\ell)$ as the fraction of errors of weight $j$ with $i$ Pauli $\M{Z}$ and $\ell$ Pauli $\M{X}$ errors that are not corrected (thereby leading to a logical error).  
Then, we can write
\begin{align}
\label{eq:rho_L}
    \rho_L = \sum_{j = 1}^{n} \binom{n}{j}(1-\rho)^{n-j} \rho^j (1-\beta_j)
\end{align}
where
\begin{align}
    \nonumber &1-\beta_j = \\ 
    &\frac{1}{\rho^j}\sum_{i = 0}^{j}\binom{j}{i} \, \pz^i \, \sum_{\ell = 0}^{j-i} \binom{j-i}{\ell}\, \px^\ell \, \py^{j-i-\ell} f_j(i,\ell)\,.
\end{align}
Considering a symmetric error correcting code with bounded distance decoding which corrects up to $e_\mathrm{g} $ errors, we have that $f_j(i,\ell) = 0$ for $j \le e_\mathrm{g}$ and $f_j(i,\ell) = 1$ otherwise. In the case of an asymmetric code able to correct $e_\mathrm{g}$ generic errors plus $e_\mathrm{Z}$ Pauli $\M{Z}$ errors (see Section~\ref{sec:TheoPerfBoundedDist}), $f_j(i,\ell) = 0$ 
for $j \le e_\mathrm{g}$, $f_j(i,\ell) = 0$ for $e_\mathrm{g} < j \le e_\mathrm{g} + e_\mathrm{z}$ and $i \geq e_\mathrm{z}$, and $f_j(i,\ell) = 1$ otherwise. 
For channels with $\px = \py$ (e.g., depolarizing and asymmetric) we have that
\begin{align}
\label{eq:beta_j}
    1-\beta_j = \frac{1}{(A+2)^j}\sum_{i = 0}^{j}\, A^i\,\sum_{\ell = 0}^{j-i} \binom{j}{i} \binom{j-i}{\ell} f_j(i,\ell)\,
\end{align}\\
and $\beta_j = \beta_j(A)$. The dependence on $A$ will be indicated only when necessary. In the following we will assume $\px = \py$. 

For a symmetric code, starting from \eqref{eq:rho_L} we obtain the upper bound 
\begin{align}
\label{eq:error_probWithBetaUB}
\nonumber
\rho_\mathrm{L} 
 \leq & \left(1-\beta_{t+1}\right) \binom{n}{t+1}\rho^{t+1} \left(1-\rho\right)^{n-t-1} \\
&+ \, \sum_{j = t+2}^{n} \binom{n}{j}(1-\rho)^{n-j} \rho^j\,.
\end{align}
Also, we can approximate the logical error rate
for $\rho \ll 1$ as
\begin{align}
\label{eq:error_probWithBetaApprox}
\rho_\mathrm{L} 
&\approx \left(1-\beta_{t+1}\right) \binom{n}{t+1}\rho^{t+1} \,.
\end{align}
Note that equation \eqref{eq:error_probWithBetaApprox} differs from \eqref{eq:error_probBDasy} in that the latter assumes that all errors with weight greater than $t$ are not correctable.  
The asymptotic slope in log-log domain remains $t+1$, but with an offset depending on $(1-\beta_{t+1})$, compared to \eqref{eq:error_probBDasy}.

In a similar way, we can find the asymptotic error correction capability of an asymmetric code.  In particular, for $\rho \ll 1$ the performance of the code becomes
\begin{align}
\label{eq:betaj_asym}
    \rho_\mathrm{L} 
\approx  ~ &(1-\beta_{e_\mathrm{g}+e_\mathrm{Z}+1}) \binom{n}{e_\mathrm{g}+e_\mathrm{Z}+1} \rho^{e_\mathrm{g}+e_\mathrm{Z}+1} 
\notag \\
&+ (1-\beta_{e_\mathrm{g}+1}) \binom{n}{e_\mathrm{g}+1} \rho^{e_\mathrm{g}+1} \,.
\end{align}

As indicated by \eqref{eq:rho_L} and its approximations, to calculate the performance of quantum codes we need to determine the fraction of correctable errors $\beta_j$.  

\section{Undetectable errors weight enumerator from Quantum MacWilliams identities} \label{sec:MacWilliams_intro}

\subsection{Undetectable Error \ac{WE}}

\begin{definition}[Logical operators]
The logical operators of a $[[n,k,d]]$ \ac{QECC} are the elements of the set $\mathcal{N}(\mathcal{S}) \setminus S$, namely the operators that commute with the stabilizer but are not contained in it.
\end{definition}

\begin{definition}[Undetectable errors]
The undetectable errors operators are those coincident with the logical operators. They transform a codeword into another codeword and are therefore undetectable.
\end{definition}
Thus, the set of logical operators coincides with the set of undetectable errors. In the following, we will use the two terms interchangeably. 

The {\it undetectable errors \acl{WE}} for a $[[n,k,d]]$ quantum code is 
\begin{align}
\label{eq:deflogicals_eq}
L(z) &= \sum_{w = 0}^{n} L_w z^w
\end{align} 
where $L_w$ is the number of undetectable errors (logical operators) of weight $w$. 

\label{thm:logicals} 
Then, we will show that $L(z)$ can be written as 
\begin{align}
\label{logicals_eq}
L(z) &= \frac{1}{2^k} B(z) - \frac{1}{4^k} A(z) 
\end{align} 
where 
$A(z) = \sum_{w = 0}^{n} A_w z^w$, $ B(z) = \sum_{w = 0}^{n} B_w z^w$, 
\begin{equation}\label{eq:Adtheo}
    A_w=4^k \sum_{\M{E}_w}\left|\M{E}_w \cap \mathcal{S}\right|
\end{equation}
\begin{align}
\label{eq:MacWilliams_coeff}
B_w &= \frac{1}{2^n} \sum_{\ell = 0}^{n} \; \sum_{s = 0}^{w} \;\; \binom{\ell}{s} \binom{n-\ell}{w-s} (-1)^s 3^{w-s}A_\ell
\end{align}
and where the sum in \eqref{eq:Adtheo} is over all operators $\M{E}_w \in \mathcal{G}_n$ of weight $w$. 

We prove now that \eqref{logicals_eq}, together with  \eqref{eq:Adtheo} and \eqref{eq:MacWilliams_coeff}, gives indeed the undetectable \ac{WE}. To this aim, let us consider the operators $\M{E}_w \in \mathcal{G}_n$ with weight $w$, i.e., containing exactly $w$ Pauli operators different from the identity.
For any two hermitian operators $\M{O}_1$ and $\M{O}_2$ we can introduce two \acp{WE} $A_w$ and $B_w$ 
\cite{ ShoPetLaf:97, Rai:98} 
\begin{align}
\label{eq:MacWilliams_Ad}
A_w(\M{O}_1,\M{O}_2) &=  \sum_{  \M{E}_w} \tr \left(\M{E}_w   \M{O}_1 \right) \tr  \left(\M{E}_w \M{O}_2 \right)\\
\label{eq:MacWilliams_Bd}
B_w(\M{O}_1,\M{O}_2) &= \sum_{ \M{E}_w} \tr \left(\M{E}_w   \M{O}_1 \M{E}_w \M{O}_2 \right)  
\end{align}
where the sum is over all the $\M{E}_w$, and $ w = 0,\ldots,n$. We will often drop the operators $\M{O}_1,\M{O}_2$ when the dependence is clear in the context.
In the case in which  $\M{O}_1 = \M{O}_2 = \Pi_{\mathcal{C}}$, the projector onto a $[[n,k,d]]$ binary stabilizer code, $A(z)$ and $B(z)$ carry some important properties of the code. Indeed, let $\M{G}_i \in \mathcal{S}$, with $i = 1,...,n-k$, generators of the stabilizer group of a code, then \cite{CaoChuLac:22}
\begin{align}
\label{eq:MacWilliamss}
\Pi_{\mathcal{C}} &= \frac{1}{2^{n-k}} \prod_{i = 1}^{n-k} \left(\M{I} + \M{
G}_i\right)  = \frac{1}{2^{n-k}}\sum_{\M{S} \in \mathcal{S}} \M{S} \,.
\end{align}
It can be shown \cite{CaoChuLac:22} that the \ac{WE} $A(z)$ defined in \eqref{eq:MacWilliams_Ad} is proportional to the stabilizer \ac{WE} 
\begin{align}
\label{eq:MacWilliams_stab}
    \frac{1}{4^k} A(z) &= \sum_{w = 0}^{n}\sum_{\M{E}_w} \left|\M{E}_w \cap \mathcal{S}\right| z^w.
\end{align}
Moreover, $B(z)$ is proportional to the normalizer \ac{WE} \cite{CaoChuLac:22}
\begin{align}
\label{eq:MacWilliams_norm}
    \frac{1}{2^k} B(z) &= \sum_{w = 0}^{n}\sum_{\M{E}_w} \left|\M{E}_w  \cap \mathcal{N(S)} \right| z^w.
\end{align}
In order to find the relation between $A_w$ and $B_w$ we write the associated \acp{WE} in the form
\begin{align}
\label{eq:MacWilliams_Az2}
\nonumber A(v,z) &= \sum_{w = 0}^{n} A_w v^{n-w} z^w \\ \quad B(v,z) &= \sum_{w = 0}^{n} B_w v^{n-w} z^w\,.
\end{align} 
These two polynomials are related through the quantum MacWilliam identities \cite{MacSlo:77,Rai:98, CaoChuLac:22}
\begin{align}
\label{eq:MacWilliams_id}
B(v,z) &=    A\left(\frac{v+3z}{2} , \frac{v-z}{2}\right) .
\end{align}
Using \eqref{eq:MacWilliams_id} in \eqref{eq:MacWilliams_Az2} we get \eqref{eq:MacWilliams_coeff}. 
Finally, the number of undetectable errors of weight $w$ is given by the number of operators of weight $w$ which commute with $\mathcal{S}$, given in \eqref{eq:MacWilliams_norm}, minus the number of stabilizers of weight $w$, given by \eqref{eq:MacWilliams_stab}, which leads to \eqref{logicals_eq}.

\smallskip
The evaluation of the undetectable errors weight enumerator \eqref{logicals_eq} requires therefore only computing the $A_w$ in \eqref{eq:Adtheo}, which  can be carried out by direct inspection of $\mathcal{S}$ for small code sizes, or more in general by using the tools for the computation of the weight distribution of classical codes as  discussed in Section~\ref{sec:MacWilliams_B2}. 

For example, let's consider the [[3,1]] repetition code, able to correct one bit flip error. In this case, we have
%
$\arraycolsep=2.8pt
\begin{array}{lllllll} 
\M{G}_1 = \mathbf{Z}_1 \mathbf{Z}_2 , &
\M{G}_2 = \mathbf{Z}_2 \mathbf{Z}_3. & 
\end{array} $ 
%
Considering $\M{O}_1 =~ \M{O}_2 =~ \Pi_{\mathcal{C}}$, we have from \eqref{eq:MacWilliamss} that the stabilizer is $\mathcal{S} = \{ \mathbf{I}_1 \mathbf{I}_2\mathbf{I}_3, \mathbf{Z}_1 \mathbf{Z}_2, \mathbf{Z}_2 \mathbf{Z}_3, \mathbf{Z}_1 \mathbf{Z}_3 \}$. If we combine \eqref{eq:MacWilliamss} and \eqref{eq:MacWilliams_Ad}, we obtain $\frac{1}{4}A(z) = 1 + 3 z^2$. Then we use \eqref{eq:MacWilliams_coeff} to compute $\frac{1}{2}B(z) = 1 + 3 z + 3 z^2 + 9 z^3$. Finally, the undetectable errors \ac{WE} is: $L(z) = 3 z + 9 z^3$.

More in general, a trivial way to obtain $A(z)$ for the $[[n,k,d]]$ code that we want to study is to  compute all  linear combinations among the set of generators. Alternatively, it is possible to consider the connection between stabilizer codes and codes over the Galois field $GF(4)$ by identifying the operators $\M{I}$, $\M{X}$, $\M{Z}$ and $\M{Y}$ with the four elements of the field  \cite{CalRaiSho:98, Got:09}. Hence, the evaluation of $A(z)$ can be seen as the computation of the weight distribution of classical codes over $GF(4)$. 
Although this problem may be classified as NP-hard  \cite{BerElwMce:78}, a variety of advanced and optimized algorithms have been developed. These algorithms surpass traditional brute force methods in efficiently calculating key metrics such as the weight distribution and the minimum weight. 
Some of them are the Brouwer–Zimmermann algorithm and its various modifications for cyclic codes, quasi-cyclic codes, and divisible codes \cite{GulAarBha:99, BouIliBou:21, Zim:96, Bet:06, Can:98, Gra:06, Lio:16}. Such algorithms are implemented in software tools related to coding theory, such as MAGMA \cite{BosWieCan:97}. For all codes analyzed in this paper, the required time for the computation on a laptop of the weight distribution was less then one second. 
Recently, new techniques have been developed for computing the quantum weight enumerator polynomial $A(z)$, which, for some degenerate stabilizer codes, provide up to an exponential speed up compared to the previous methods \cite{Cao:23}. 

\subsection{Bounds on $
\beta_{t+1}$ via Undetectable Errors Weight Enumerator} 
\label{sec:MacWilliams_B2A}
The performance of an $[[n,k,d]]$ \ac{QECC} is mainly determined, according to \eqref{eq:error_probWithBetaUB}, \eqref{eq:error_probWithBetaApprox}, and \eqref{eq:betaj_asym}, by the value of $\beta_{t+1}$. We show here that, even without analyzing in details the logical operators of a code, it is possible to exploit the undetectable error \ac{WE} $L(z)$ in order to find upper bounds on the performance for a depolarizing channel. Specifically, considering the general case of a complete \ac{MW} decoder, we will derive several lower bounds on the value of $\beta_{t+1}$, indicated as $\hat{\beta}_{t+1} \leq \beta_{t+1}$, for some families of codes. Unless otherwise stated we will assume $d$ odd. 

A first bound, valid in general for stabilizer codes, is obtained assuming that each logical operator of weight $w$ can be caused by all the $3^{t+1}$ different Pauli errors of weight $t+1$, as
\begin{align}
\label{mac_beta_est}
    \hat{\beta}_{t+1} = \left (   1 - \frac{3^{t+1} \, \sum_{w = 2t + 1}^{2t + 2}L_w \, \binom{w}{t+1} }{\binom{n}{t+1}\,3^{t+1}} \,\, \right)^{+}  \,.
\end{align}
In this equation, $L_w$ is the number of logical operators of weight $w$, $3^{t+1}$ is the number of permutations of the different Pauli operators of weight $t+1$ that cause a logical operator, and $\binom{w}{t+1}$ refers to the number of patterns of errors (made of fixed Pauli operators) with weight $t+1$ that can realize the corresponding logical operator. 
For instance, if we consider $t+1 = 2$, we have $3^2=9$ different Pauli errors: $\M{Z}$$\M{Z}$, $\M{X}$$\M{Z}$, $\M{Z}$$\M{X}$, $\M{X}$$\M{X}$, $\M{Z}$$\M{Y}$, $\M{Y}$$\M{Z}$, $\M{Y}$$\M{Y}$, $\M{Y}$$\M{X}$, $\M{X}$$\M{Y}$. 
Moreover, a logical operator of weight $w=3$ can be produced by $\binom{3}{2} = 3$ patterns of each of the previous Pauli errors of weight $t+1$.
Note that, the summation in \eqref{mac_beta_est} starts from $2t+1$ since $L_w = 0$ for $w \leq 2t + 1$, and is limited to $2t+2$ because, when an error of weight $t+1$ is introduced by the channel, a \ac{MW} decoder will never choose a codeword which differs in more than $t+1$ positions from the original one. 
The previous bound can be made more tight if we deal with \ac{CSS} codes, where the $\M{X}$ and $\M{Z}$ corrections are performed independently. In this case the logical operators of weight $d$ are formed by one Pauli type only. Consequently, a logical operator may be caused by only two types of Pauli errors (instead of the previously considered three). For instance, let us consider the logical operator $\M{Z}$$\M{Z}$$\M{Z}$: this can be caused only by $\M{Z}$$\M{Z}$, $\M{Z}$$\M{Y}$, $\M{Y}$$\M{Z}$ and $\M{Y}$$\M{Y}$ Pauli errors.  
Therefore, for \ac{CSS}, the number of permutations of Pauli errors that cause the generic logical operator is $2^{t+1}$.
Moreover, since, for a \ac{MW} decoder, the error recovery operator has always a weight lower or equal to the weight of the channel error pattern, only half of the error patterns of weight $t+1$ cause a logical operator of weight $w=2t+2$. Hence, the fraction of corrected errors can be bounded by
\begin{align}
\label{mac_beta_est_mwpm}
    \nonumber &\hat{\beta}^\mathrm{CSS}_{t+1} = \\  & \left ( \! 1 - \frac{2^{t+1} \! \sum_{w = 2t + 1}^{2t+2}L_w \, \binom{w}{t+1} / \!  \left \lfloor \frac{w}{t+1}\right \rfloor }{\binom{n}{t+1}3^{t+1}} \right )^{+}.
\end{align}
We remark that this expression applies also to surface codes with \ac{MWPM} decoder. 
Furthermore, it is possible to obtain a tighter bound if the generators are composed by $\M{X}$ or $\M{Z}$ Pauli measurements on the same qubits. In particular, this condition holds for the category of \ac{CSS} Dual-Containing codes \cite{MacMitMcF:04}. These codes have a third of the logical operators of minimum weight composed by only $\M{Y}$ Pauli operators. Hence, we know  that these logical operators can be caused only by Pauli errors composed by $\M{Y}$ operators. In this situation, \eqref{mac_beta_est_mwpm} can be rewritten as 
\begin{align}
\label{mac_beta_est3}
\nonumber
    &\hat{\beta}^\mathrm{CSS-DC}_{t+1} =  \\
    &\left(\!\!1 \! - \! \frac{\frac{1}{3}\sum_{w = 2t +1}^{2t +2} \! L_w \!\! \,\left(2  \left(2^{t+1} - 1\right) + 1 \right) \,\!\! \binom{w}{t+1} / \!  \left \lfloor \! \frac{w}{t+1} \! \right \rfloor  }{\binom{n}{t+1}\,3^{t+1}} \right)^{\!\!+} \!\!.
\end{align}
A more detailed discussion on the derivation of this expression can be found in Appendix. 

The values provided by \eqref{mac_beta_est},  \eqref{mac_beta_est_mwpm}, and \eqref{mac_beta_est3}, can be used in \eqref{eq:error_probWithBetaUB}, \eqref{eq:error_probWithBetaApprox}, and \eqref{eq:betaj_asym} to compute upper bounds on the error rate. 
These new bounds are easy to compute, as they just need the \ac{WE} polynomial $L(z)$ derived from the MacWilliams identities. 

\emph{Remark on degeneracy:} If we want to have a more precise estimation of the error performance we should get tighter bounds on $\beta_{t+1}$, and this is only possible by a closer analysis of the code.
In particular, we should take into account the code degeneracy, which requires a more detailed description of the code logical operators, as discussed in Section~\ref{sec:MacWilliams_B2A}.   
To explain the role of degeneracy, assume we have a code with logical operators that share the same Pauli operators on $t+1$ common qubits. In the event of an error composed exclusively of these Pauli operators, a deterministic decoder will only trigger one of these logical operators. 
Therefore, by having knowledge of the structure of the logical operators within a stabilizer code, we could improve the previous bounds.
For example, the estimation $\hat{\beta}^\mathrm{CSS}_{t+1}$ of \eqref{mac_beta_est_mwpm} can be extended to account for degeneracy as 
\begin{align}
\label{mac_beta_est_deg}
    \hat{\beta}^\mathrm{CSS}_{t+1} =  \left ( 1 - \frac{2^{t+1} \, \sum_{w = 2t +1}^{2t+2}L_w \, \binom{w}{t+1} \, \gamma_w / \!  \left \lfloor \! \frac{w}{t+1} \! \right \rfloor }{\binom{n}{t+1}\,3^{t+1}} \,\,  \right )^{+}
\end{align}
where $\gamma_w \in \left[{1}/{\binom{w}{t+1}},1\right]$ is the average fraction of potential Pauli error patterns of weight $t+1$ that are not shared between two or more logical operators of weight $w$. 
Note that this parameter is contingent on both the weight and the distinct Pauli operators that constitute the logical operators.
For instance, if the code is asymmetrically degenerate with respect to the different Pauli operators, different logical operators of the same weight could share a different number of Pauli operators.  
We remark that the bounds \eqref{mac_beta_est} to \eqref{mac_beta_est3} are already in closed form and do not necessitate the evaluation of $\gamma_w$. Their computational complexity solely arises from evaluating the weight enumerator of the specific quantum code. 

\emph{Remark on even code distance:} The expressions derived in \eqref{mac_beta_est} to \eqref{mac_beta_est3} can be easily modified to the case where $d$ is even by keeping only the weight $w = 2t + 2 $ in the summation, and avoiding the division by $  \left \lfloor  \frac{w}{t+1} \right \rfloor$.
For example, for codes with even distance $d=2t+2$,  $\eqref{mac_beta_est_mwpm}$ becomes
\begin{align}
\label{mac_beta_est_mwpm_es}
     \hat{\beta}^\mathrm{CSS}_{t+1} =   \left (  1 - \frac{2^{t+1} \,\, L_{2t+2} \, \binom{2t+2}{t+1} }{\binom{n}{t+1}3^{t+1}} \right )^{+}.
\end{align}

\medskip

Let us provide now examples 
for some important quantum codes. 
As regards the $[[7,1,3]]$ Steane code\cite{SteAnd:96}, using \eqref{logicals_eq} we compute the \ac{WE} as $L(z) = 21 z^3 + 126 z^5 + 45 z^7$. Hence, by applying \eqref{mac_beta_est3} it is possible to compute $\hat{\beta_2}$ as
\begin{align}
\label{mac_beta_estSt}
    \hat{\beta}^\mathrm{CSS-DC}_2 &=  \left ( 1 - \frac{ \frac{2}{3} \, 21 \cdot 3 \cdot 3 + \frac{1}{3} \, 21 \cdot 3}{\binom{7}{2}\,3^2} \,\,  \right )^{+} \nonumber \\
    &= \frac{2}{9} \simeq 0.22 
\end{align}
i.e., at least $22\%$ of the errors of weight $j=2$ are corrected by a minimum weight decoder. For this particular code this estimate is exact, i.e.,  $\hat{\beta}_2 = \beta_2$, as will be shown in section \ref{sec:MacWilliams_B2}.
This is because the logical operators of weight $w=3$ share only $t = 1$ Pauli operators on common qubits. Hence, the code degeneracy does not affect the computation of $\hat{\beta_2}$. 

In case of the $[[9,1,3]]$ Shor code \cite{Sho:95}, from \eqref{logicals_eq} the \ac{WE} results $L(z) = 39 z^3 + 208 z^5 + 332 z^7 + 189 z^9$. Moreover, using \eqref{mac_beta_est_mwpm} we get $\hat{\beta_2} = 0$. This result, which gives us no more information with respect to the bounded distance performance, is due to the strong degeneracy of the code. 
In fact, even if the number of logical operators of weight $w = 3$ is quite large for a 9 qubits code, a lot of them share $t~+~1~=~2$ Pauli operators, affecting the actual $\beta_2$ (i.e. $\gamma_3 \ll 1$). An accurate estimation taking into account the effect of degeneracy will be provided in Section~\ref{sec:MacWilliams_B2A} by a counting argument on the logical operators.

Taking into consideration the $[[13,1,3]]$ surface code, from \eqref{logicals_eq} we first find $L(z) = 6 z^3 + 24 z^4 + 75 z^5 + 240 z^6 + 648 z^7 + 1440 z^8 + 2538z^9 + 3216z^{10} + 2634z^{11} + 1224z^{12} + 243z^{13}$. Then, considering a \ac{MWPM} decoder, from \eqref{mac_beta_est_mwpm} we have
\begin{align}
\label{mac_beta_est_mwpm13}
    \nonumber \hat{\beta}^\mathrm{CSS}_2 &= \left ( 1 - \frac{2^2 \, \left( 6\, \binom{3}{2} + 24 \, \binom{4}{2}/2 \right)} {\binom{13}{2}\,3^2} \,\,  \right )^{+} \\
    &= \frac{57}{117} \simeq 0.49
\end{align}
so that at least $49\%$ of the errors of weight $j=2$ is corrected.

The method works also for asymmetric \ac{CSS} codes. For instance, taking the $[[23,1,3/5]]$ surface code, we have $L(z) = 5 z^3 + 20 z ^ 4 + 51 z ^ 5 + 172 z ^ 6 + ...$ . Considering that errors of weight $j = 2$ can be caused only by $\M{X}$ operators, while errors of weight $j=3$ can be caused both by $\M{X}$ and $\M{Z}$ operators, from \eqref{mac_beta_est_mwpm} we obtain
\begin{align}
\label{mac_beta_est_mwpm23A}
    \nonumber \hat{\beta}^\mathrm{CSS}_2 &= \left ( 1 - \frac{2^2 \big (5 \, \binom{3}{2} + 20 \,\binom{4}{2} / {2} \big ) } {\binom{23}{2}\,3^2} \,\,  \right )^{+} \\
    &= \frac{659}{759} \simeq 0.87 
\end{align}
and
\begin{align}
\label{mac_beta_est_mwpm23B}
    \nonumber \hat{\beta}^\mathrm{CSS}_3 &= \left ( 1 - \frac{2^3 \big (51 \, \binom{5}{3} + 172 \, \binom{6}{3}/2 \big ) } {\binom{23}{3}\,3^3} \,\,  \right )^{+} \\
     &= \frac{17840}{47817} \simeq 0.63 \,.
\end{align}

The exact values of $\beta_{t+1}$ for these codes are provided below. In literature, some lower bounds on the logical error rate of surface codes have been proposed for $\rho \ll 1$, not for the depolarizing but for the phase flip or the bit flip channels \cite{Lee:21, Wat:14}. These bounds are quite far from the true performance, as they consider only channel errors leading to logical operators of minimum weight $w = 2t + 1$, i.e. straight horizontal or vertical chains (see Section~\ref{sec:NumRes}).

\subsection{Closed form expression for the \ac{WE} of Surface Codes}\label{sec:betasurface}

\begin{figure*}[t]
	\centering
\includegraphics[width=0.9\textwidth]{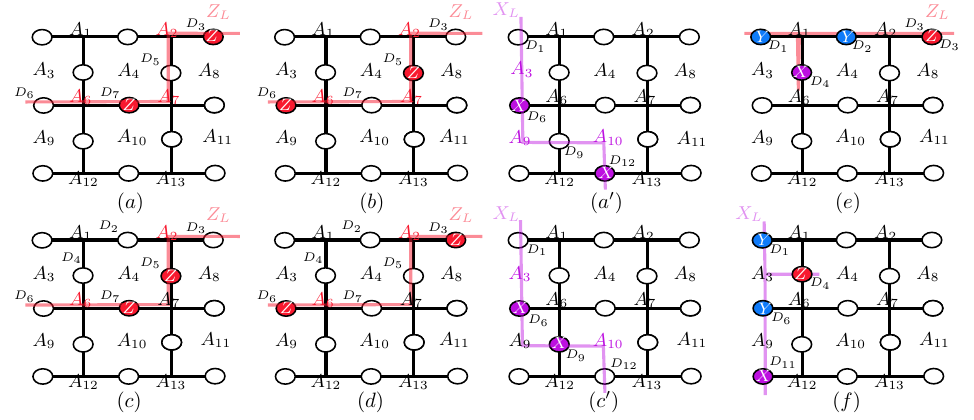}
	\caption{Example of errors leading to a logical operator of $w=4$ for the $[[ 13,1,3 ]]$ surface code. $\M{Z}$, $\M{X}$,  and $\M{Y}$ errors on qubits are depicted in red, purple, and blue, respectively.   (a) $\M{Z}_L$ occurs if $\M{Z}$ correction operators are applied on data qubits $D_5$ and $D_6$. The errors are corrected if the \ac{MWPM} decoder applies $\M{Z}$ on $D_3$ and $D_7$. (b) $\M{Z}_L$ occurs if $\M{Z}$ correction operators are applied on $D_3$ and $D_7$. The errors are corrected if the \ac{MWPM} decoder applies $\M{Z}$ on $D_5$ and $D_6$. (c) $\M{Z}_L$ occurs if $\M{Z}$ correction operators are applied on $D_3$ and $D_6$. The \ac{MWPM} decoder could apply also $\M{Z}$ operators on $D_5$ and $D_7$, correcting the error, or on $D_2$ and $D_4$, correcting the error. (d) $\M{Z}_L$ occurs if $\M{Z}$ correction operators are applied on $D_5$ and $D_7$. The \ac{MWPM} decoder could apply also $\M{Z}$ operators on $D_3$ and $D_6$ or on $D_2$ and $D_4$, causing a different logical operator. ($a^\prime, c^\prime $) Analogous examples for $\M{X}_L$ logical operator. ($e, f $) Logical operators of $w = 4$ caused by $\M{Y}\M{Y}$ errors.}
	\label{Fig:surface_logicals}
\end{figure*}

As evident from \eqref{mac_beta_est_mwpm}, when dealing with surface codes with \ac{MWPM}, a channel error with weight $j = t + 1$ has the potential to induce logical operators of weight $w = 2t + 1$ and $w = 2t + 2$, when $d$ is odd. On the other hand, when $d$ is even $\beta_{t+1}$ requires only the knowledge of the number of logical operators of weight $w = 2t+2$.
Consequently, in order to assess the performance bounds of surface codes over a depolarizing channel, it suffices to determine the \ac{WE} coefficients associated with these two degrees. 
In the following, we present an expression for determining these values without the necessity of computing the MacWilliams identities. 
Specifically, given a surface code with minimum distance $d$, there are exactly $2 d$ logical operators of weight $w = d$. These operators correspond to straight horizontal $\M{Z}$ and vertical $\M{X}$ chains crossing the lattice, as depicted in Fig.~\ref{Fig:surface}. 
In fact, all other paths from a boundary to an opposite one have at least $d + 1$ Pauli operators. We conclude that $L_{d} = 2d$. 
Furthermore, logical operators of weight $w = d + 1$ can be classified into two distinct groups. The first category, composed by $4 (d-1)^2 $ operators, representing the chains starting from a boundary and reaching the opposite side of the lattice with only one turn.
Some examples of these $\M{Z}_L$ and $\M{X}_L$ operators for the $[[13,1,3]]$ surface code are depicted in Fig.~\ref{Fig:surface_logicals} $(a)$ and $(a^\prime)$. 
The last group of logical operators is a tricky one. 
In particular, these operators are obtained from the $\M{Z}_L$ ($\M{X}_L$) operators of weight $w = d$ by applying one site (plaquette) generator. 
Let us consider firstly a logical operator with $w = d$ that traverse qubits which are measured by four-qubits generators. 
If we apply a four-qubits generator to these logical operators we end up with another logical operator of weight $ w = d + 2$. 
For this reason, we exclude these from the counting.
On the other hand, considering logical operators running along a boundary of the lattice, we can apply a three-qubits generator to end up with another logical operator of weight $w = d + 1$, corresponding to the overall count of three-qubits generators, i.e., $4(d-1)$. 
An example of these operators is shown in Fig.~\ref{Fig:surface_logicals} $(e)$ and $(f)$.
Counting all the combinations, we conclude that $L_{d+1} = 4 d(d-1) $. 

In summary, a $[[d^2+(d-1)^2,1,d]]$ surface code has \ac{WE} 
\begin{align}
\label{eq:Ldsurf}
    L_d = 2d \qquad L_{d+1} = 4 d (d-1) 
\end{align}
which can be used in \eqref{mac_beta_est_mwpm} and \eqref{mac_beta_est_mwpm_es}. 
For example, for the $[[41,1,5]]$ surface code $L_5 = 10$ and $L_6 = 80$, resulting in $\hat{\beta}_3 = 0.975$. 

As a check, \eqref{eq:Ldsurf} is in accordance with what we can obtain by applying \eqref{logicals_eq} and \eqref{eq:MacWilliams_coeff} to the numerically computed $A(z)$ for $d=10$ in \cite{Cao:23}. 
Therefore, due to \eqref{eq:Ldsurf} we do not need to compute $A(z)$ for evaluating the performance of surface codes (see also Section~\ref{sec:NumRes}).  

\subsection{ \ac{QECC} Performance via Logical Operators Analysis}\label{sec:MacWilliams_B2}

Now we show that it is possible to obtain the exact $\beta_j$ parameters with no need of simulations. This requires an analysis of the structure of the code logical operators, which we do explicitly for some small-size codes. The same approach may become too complicated for large codes, where the use of the closed-form bounds \eqref{mac_beta_est_mwpm}, and \eqref{mac_beta_est3} is preferable. 
Since we are interested in the value of $\beta_j$, we have to find the fraction of errors of weight $j$ that can cause the decoder to miscorrect and induce a logical error.
In order to compute it, we will consider not only $L(z)$ but also the structure of the stabilizer.

In the case of a \ac{CSS} stabilizer code, taking into account that the total number of different error patterns of weight $j$ is $\binom{n}{j}$, we can express $\beta_j$ as
\begin{align}
\label{mac_beta}
    \nonumber &\beta_j = \\
    &1 - \frac{\sum_{i=0}^j \, A^i \sum_{\ell=0}^{j-i}\, \sum_{w} L_w(i,\ell) \, \mu^{(w)}_j(i,\ell) \, \gamma_w(i,\ell) }{(A+2)^j {\binom{n}{j}}}
\end{align}
where: $L_w(i,\ell)$ stands for the number of logical operators of weight $w$ that can be caused by the combined action of channel errors of weight $j$ composed by $i$ $ \M{Z}$ and $ \ell$ $\M{X}$ Pauli operators and $w-j$ correction operators applied by a complete decoder; $\mu^{(w)}_j(i,\ell)$ refers to the number of different patterns of errors with weight $j$ that can induce the corresponding logical operator of weight $w$; and $\gamma_w(i,\ell) \in \left[{1}/{\binom{w}{j}},1\right]$ is the average fraction of potential Pauli error patterns of weight $j$ that are not shared between two or more logical operators. 
Note that \eqref{mac_beta} and \eqref{eq:beta_j} are equivalent, since
\begin{align}
\label{Ctot2}
   \nonumber \sum_{w} L_w(i,\ell) \, &\mu^{(w)}_j(i,\ell) \, \gamma_w(i,\ell) = \\ 
   & \quad \binom{n}{j} \binom{j}{i} \binom{j-i}{\ell} f_j(i,\ell).
\end{align}
In particular, on both sides of \eqref{Ctot2} we find the total number of logical errors induced by the weight $j$ error class identified by $i$ operators of type $ \M{Z}$ and $\ell$ operators of type $\M{X}$.

In the following, we provide some examples over a depolarizing channel ($A=1$), in order to clarify the reasoning behind the evaluation of $\beta_j$. 

First, let us consider the $[[5,1,3]]$ perfect code \cite{LafRayMiq:96}. Note that this is not a \ac{CSS} code. Using \eqref{logicals_eq} we obtain $L(z) = 30 z^3 + 18 z^5$, so we know that the number of logical operators of weight $3$ is $30$. Then, we have 
\begin{align}
    \beta_2 = 1 - \frac{30\cdot3}{\binom{5}{2} 3^2 } = 0
\end{align}
where we have considered that the total number of pairs of Pauli errors is $\binom{5}{2} 3^2$, and that each logical operator of weight $3$ can be caused exactly by three different combinations of errors of weight $j = 2$. This result was expected since the code is perfect. 

A more interesting case is that of the $[[7,1,3]]$ Steane code\cite{SteAnd:96}. 
Starting from $L(z) = 21 z^3 + 126 z^5 + 45 z^7$ and assuming a \ac{MW} decoder, we calculate 
\begin{align}
\label{eq:Beta2Steane}
\notag
    \beta_2 = ~ &1 - \frac{1}{\binom{7}{2} 3^2 } \bigg[7 \mu^{(3)}_2(2,0) + 7 \mu^{(3)}_2(1,0) \\ \notag & \quad + 7 \mu^{(3)}_2(0,2) 
      + 7  \mu^{(3)}_2(0, 1) + 7  \mu^{(3)}_2(0, 0) \bigg]  \\
    = ~ & 1 - \frac{7\, (3 + 6) + 7\, (3 + 6) + 7 \cdot 3   }{\binom{7}{2} 3^2 } =\frac{2}{9} \simeq 0.22.
\end{align}
 To derive this value we observed that, unlike the perfect code, each logical operator of $w = 3$ is composed of only one kind of Pauli operator. Specifically, we have seven  $\M{X}$$\M{X}$$\M{X}$, seven $\M{Z}$$\M{Z}$$\M{Z}$, and seven $\M{Y}$$\M{Y}$$\M{Y}$ logical operators. In addition, $\gamma_3 = 1$ since there are no logical operators that share Pauli operators on the same qubits. Note that these logical operators act on the same qubits, leading to a simple expression. Moreover, we have to consider that each one of these $\M{X}$ ($\M{Z}$) logical operators can be generated by  $\M{Z}$$\M{Z}$ ($\M{X}$$\M{X}$) errors, and also by $\M{X}\M{Y}$ ($\M{Z}\M{Y}$), since single errors can always be corrected, while $\M{Y}$ logical operators are caused only by two $\M{Y}$ errors. 
From \eqref{eq:error_probWithBetaApprox} and \eqref{eq:Beta2Steane} we then obtain for the $[[7,1,3]]$ code over the depolarizing channel
\begin{align}
\label{eq:rhoLSteane}
    \rho_L \approx 16.3 \rho^2 \,, \quad \rho \ll 1 \,.
\end{align}
 
As a third example let us analyze the $[[9,1,3]]$ Shor code \cite{Sho:95}. From \eqref{logicals_eq} the \ac{WE} results $L(z) = 39 z^3 + 208 z^5 + 332 z^7 + 189 z^9$. Assuming a \ac{MW} decoder, 39 undetectable errors of weight three are caused by channel errors of weight two. A closer look reveals that the 39 operators include three $\M{X}\M{X}\M{X}$, 27 $\M{Z}\M{Z}\M{Z}$, and 9 $\M{Y}\M{Y}\M{X}$ logical operators. 
The code is degenerate with respect to $\M{Z}$ Pauli errors, so it is necessary to treat differently the logical operators of each type. In particular, since each channel error of the kind  $\M{Z}\M{Z}$ can cause three different $\M{Z}$$\M{Z}\M{Z}$ logical operators,  $\gamma_3(2,0) = \gamma_3(1,0) = \frac{1}{3}$. For instance, the channel error $\M{Z}_1$$\M{Y}_4$ could cause either $\M{Z}_1$$\M{Z}_4$$\M{Z}_7$, $\M{Z}_1$$\M{Z}_4$$\M{Z}_8$, or $\M{Z}_1$$\M{Z}_4$$\M{Z}_9$. (The error $\M{X}_4$ is corrected by the decoder).

Thus, considering that  $\gamma_3(0,0) = \gamma_3(0,1) = \gamma_3(0,2) = 1$, we obtain   
{
\begin{align}
\label{eq:Beta2Shor}
\notag 
 \beta_2 &= 1 - \frac{1}{\binom{9}{2} 3^2} \bigg[27\,\,[ \mu^{(3)}_2(2,0) +  \mu^{(3)}_2(1,0)]\,\, \gamma_3(2,0) \\ \notag &+ 3 \, [\mu^{(3)}_2(0,2) + \mu^{(3)}_2(0,1)]+ (27+9) \, \mu^{(3)}_2(0,0) \bigg] \\ \notag
    &= 1 - \frac{27 \left( 3 +  6\right) \frac{1}{3} + 3 \cdot 3 + 3 \cdot 6 + 27+9  }{\binom{9}{2} 3^2 } \\ 
    &= \frac{5}{9} \simeq 0.56
\end{align}
} 
\noindent From \eqref{eq:error_probWithBetaApprox} and \eqref{eq:Beta2Shor} we obtain for the $[[9,1,3]]$ code
\begin{align}
\label{eq:rhoLShor}
    \rho_L \approx 16 \rho^2 \,, \quad \rho \ll 1 \,.
\end{align}

Finally, we investigate the $\beta_j$ values for surface codes, which belong to the class of \ac{CSS} codes, assuming \ac{MWPM} decoding. 
Considering the $[[13,1,3]]$ surface code, we have $L(z) = 6 z^3 + 24 z^4 + 75 z^5 + ...$. As before, we need to analyze how channel errors of weight $j=2$ cause logical operators. 
Let us first look at the six logical operators of weight three. For the surface codes we know that the logical operators cross the lattice from side to side. Thus, there are three $\M{X}\M{X}\M{X}$ (crossing horizontally) and three $\M{Z}\M{Z}\M{Z}$ (crossing vertically) logical operators. As a result, $\mu^{(3)}_2(0,2) = \mu^{(3)}_2(2,0) = 3$, given that, for instance, a $\M{Z}\M{Z}\M{Z}$ logical operator can arise from $\binom{3}{2}$ error patterns of the $\M{Z}\M{Z}$ type. Furthermore, $\mu^{(3)}_2(0,1) = \mu^{(3)}_2(1,0) = 6$, as a $\M{Z}\M{Z}\M{Z}$ logical operator is induced by $2\binom{3}{2}$ error patterns of the $\M{Z}\M{Y}$ kind. 
In this code we have also logical operators of weight $w = 4$, as illustrated in Fig.~\ref{Fig:surface_logicals}. 
Specifically, among the $L_4 =24$ logical operators, 16 are composed by $\M{X}\M{X}\M{X}\M{X}$ and $\M{Z}\M{Z}\M{Z}\M{Z}$, and the remaining eight are of type $\M{Y}\M{Y}\M{X}\M{Z}$, as illustrated in Fig.~\ref{Fig:surface_logicals} $(e), (f)$. 
For these eight cases, we are left, after \ac{MWPM} decoding, with a logical operator with three $\M{Z}$ or three $\M{X}$, which have been  already counted when discussing the weight $w = 3$.  
As regards the other 16 logical operators, we have $\mu^{(4)}_2(2,0) = \mu^{(4)}_2(0,2) = \mu^{(4)}_2(0,0) = 2$. In particular, for each logical operator of weight $w = 4$ there are $\binom{w}{j} = \binom{4}{2} = 6$ different patterns of errors of weight $j = 2$ that can cause it. However, one of them is always corrected (due  to the degeneracy of the code), while another one cause a logical operator of weight $w = 3$ that we have already taken into consideration. About the remaining four, they cause, in pairs, the same syndrome. 
We consider a deterministic decoder, such as the \ac{MWPM}, that associate one error pattern to each syndrome. Thus, only two patterns will not be corrected. Also, note that, among the four patterns of errors of weight $j=2$ that can cause a logical operator of weight $w = 4$,  one is in common with another logical operator. For example, the error pattern $\M{Z}_3\M{Z}_6$ depicted in Fig.~\ref{Fig:surface_logicals}(d) may also cause the logical operator $\M{Z}_2\M{Z}_3\M{Z}_4\M{Z}_6$. 
Hence, among the four faulty error patterns corresponding to the pair of logical operators, one pattern is repeated twice.
Therefore we have  $\gamma_4(i,\ell) = {3}/{4}$. 
In Tab.~\ref{tab:param}, we report, for the $[[13,1,3]]$ surface code, the values of $L(z)$, $\gamma_w(i,\ell)$, and $\mu_j^{(w)}(i,\ell)$ that are needed in order to compute $\beta_2$. If we put these parameters into \eqref{mac_beta}, we obtain

\input{Figures/cute_table3bis}
\begin{align}
\label{eq:Beta2Surf}
    \beta_2 = \frac{267}{351} \simeq 0.76. 
\end{align}
From \eqref{eq:error_probWithBetaApprox} and \eqref{eq:Beta2Surf} we obtain for the $[[13,1,3]]$ surface code over the depolarizing channel
\begin{align}
\label{eq:rhoLSurf}
    \rho_L \approx 18.7 \rho^2 \,, \quad \rho \ll 1 \,.
\end{align}
Using \eqref{mac_beta} it is also possible to obtain the value of $\beta_j$ for asymmetric channels. For instance, in the case of the [[13,1,3]] surface code over a phase flip channel, we have only three $\M{Z}$ logical operators with $w=3$ and eight $\M{Z}$ logical operators with $w=4$, giving
\begin{align}
\label{eq:computeBeta2}
\nonumber
    \beta_2 &= 1 - \frac{3 \mu^{(3)}_2(0,2) + 8 \mu^{(4)}_2(0,2)  \gamma_4(0,2)}{\binom{13}{2}} \\
    &= 1 - \frac{3\cdot3 + 8\cdot 2 \frac{3}{4}}{\binom{13}{2}} 
    = \frac{19}{26} \simeq 0.73 \,.
\end{align}
\subsection{$\beta_j$ by exhaustive search}
\label{sec:exSearch}

A different approach to compute the exact $\beta_j$ coefficients for a given code and a given decoder is by exhaustive search. In Tab.~\ref{tab:Err} we report for some surface codes the percentage of non-correctable errors for each error class $f_j(i,\ell)$, which we have evaluated by exhaustive search with a \ac{MWPM} decoder. 
\input{Figures/cuteTable}
In doing so we exploited the Lemon C++ library \cite{DezBalJut:11}, which provides an efficient implementation of graphs and networks algorithms. For example, in the case of the $[[23,1,3/5]]$ surface code, it results $f_2(0,2) = 0.16$. Once we have $f_j(i,\ell)$, we can compute the value of $1-\beta_j$ for arbitrary $A$ (for example in Tab.~\ref{tab:Err} we report the cases $A=1$, $A=10$, $A=100$, and $A \to \infty$), by weighting the percentages of non-correctable errors as indicated by \eqref{eq:beta_j}. 
From Tab.~\ref{tab:Err}, we observe that, as anticipated, surface codes can correct a large number of errors above the guaranteed error correction capability.
Note that, as the code's length increases, performing an exhaustive search to derive the $\beta_j$ values may pose significant computational demands.

\section{Noisy Syndrome Measurement} \label{Sec_syndromeExtr}

In realistic quantum systems, syndrome measurements are inherently noisy and often require repeated execution over time to yield reliable outcomes. 
In this section, we shift our attention to  syndrome extraction circuits and introduce a  framework that accounts for the presence of noise during measurement.
We recall that a gadget for a specific function is defined as a circuit to perform that function on the encoded state \cite{Got:09}.
Then, we replace each measurement in the original circuit with a fault-tolerant gadget that replicates its intended action on logical qubits, ensuring it behaves as the ideal measurement would in the unencoded circuit.
By systematically propagating error probabilities through the circuit, we derive upper bounds on the resulting physical error rate.
When combined with the previously established theoretical bounds, this approach enables us to estimate the logical error rate including the effects of the noisy gadget, offering a practical tool for analyzing performance in real-world quantum error correction scenarios.
Specifically, in realistic syndrome extraction gadgets, two main challenges arise. 
First, syndrome measurements are inherently noisy, making a single-shot measurement unreliable \cite{Sho96:CatStates, Got:09}.
This issue is typically addressed by repeating the measurement until the same syndrome is obtained $t+1$ times consecutively \cite{Sho96:CatStates}.
In many practical scenarios, for distance-$d$ codes, it is common to perform $d$ repeated measurements \cite{NieChu:10, DenKitLan:02, hog24:dRep}.
The second challenge stems from error propagation during the extraction process.
In particular, errors on the syndrome qubit can spread to multiple data qubits through entangling gates, leading to high-weight errors known as hook errors. 
To mitigate these effects, techniques such as the use of cat states or flag qubits have been developed, which can signal the presence of correlated faults \cite{Sho96:CatStates, Cha18:FlagErrCorr, Cha18:FlagsDet}. 
An alternative is the Steane error correction gadget, which, by construction, implements a fault-tolerant gate through transversal operations and destructive measurement of a logical ancilla 
\cite{Ste97:SteaneGadget}. \\

In the following, we consider realistic syndrome extraction gadgets that include faulty initialization, single-qubit and two-qubit gates, and measurements. 
We assume that after each of these operations, a depolarizing channel acts on the qubits involved, with error probabilities specific to each type of gate. 
In addition to these, we also include an initial depolarizing channel acting on each data qubit before the syndrome extraction gadget is applied.
Moreover, in practical quantum experiments, error correction is executed over multiple consecutive correction cycles \cite{AchRajAle:22, ai2024quantum}.
Between cycles, data qubits may accumulate errors due to decoherence or as a result of gate operations performed during encoded computation.
Importantly, errors arising from an incorrect syndrome bit in a given cycle, as well as those introduced by the final two-qubit gate during syndrome extraction (which are not detected within the same cycle), can be dealt with in subsequent cycles. 

\subsection{Cat States}

In this approach, the stabilizer generator $\M{G}_i$ of weight $\gamma_i$ is measured using a cat state (or GHZ state) composed of $\gamma_i$ ancilla qubits.
This configuration ensures that each controlled gate in the extraction circuit interacts with a dedicated syndrome qubit, thereby limiting the spread of errors \cite{Sho96:CatStates}. 
The entire syndrome extraction procedure is assumed to be repeated for $r > t$ shots \cite{Got:09, Sho96:CatStates}.
The error probability of a specific syndrome bit can be bounded by considering that $t+1$ consecutive syndrome extractions providing the same syndrome, i.e., no errors have occurred~\cite{Sho96:CatStates}.
This~is
\begin{align}
\label{eq:depoCatMeas}
    P_{i} \le 1 - (1 - p_{i})^{t + 1}
\end{align}
where $p_{i}$ is the error probability in a single shot syndrome bit measurement. 
This can be upper bounded as
\begin{align}
\label{eq:depoCatMeas2}
    \notag p_{i} \leq 1 - &(1 - \rho_\mathrm{init}^\mathrm{cat}) \Big[(1 - \rho_{1Q}) \\
    &\times (1 - \rho_{2Q})(1 - \rho_{\mathrm{meas}})\Big]^{\gamma_i}
\end{align}
with $\rho_\mathrm{init}^\mathrm{cat}$, $\rho_{1Q}$, $\rho_{2Q}$, and $\rho_{\mathrm{meas}}$ denote the depolarizing error probabilities associated with the cat state initialization, single-qubit gates, two-qubit gates, and qubit measurement, respectively.
Moreover, note that a \ac{MW} decoder can apply corrections to at most $t$ qubits.
As a result, a single syndrome bit error can, in the worst case, be equivalent to a Pauli error affecting $t$ qubits.
For a given stabilizer generator, the probability that a physical qubit is affected due to a faulty syndrome bit is $t / n$, where $n$ is the total number of qubits.
Therefore, the overall qubit error probability resulting from faulty syndrome measurements can be upper bounded as
\begin{align}
\label{eq:MEasERRCAT}
    P_\mathrm{syn} \leq \frac{t}{n} \left [ 1 - \prod_{i = 1}^{n - k}(1 - P_{i}) \right].
\end{align}
To compute the depolarizing error probability after syndrome extraction, we also account for an initial depolarizing channel on each data qubit, as well as a depolarizing channel associated with each two-qubit gate that interacts with a data qubit.
Since we aim to obtain an upper bound for the depolarizing error probability, which is assumed to be the same for every data qubit, we are interested in the maximum number of two-qubit gates applied to any single qubit. 
Thus, we define $\delta_\mathrm{max} = \underset{j}{\max} \, \delta_j$, where $\delta_j$ represents the number of two-qubit gates applied to qubit $j$ in a single round of syndrome extraction.
Hence, the equivalent depolarizing error probability on each qubit at the end of the measurement based on cat state can be upper bounded as
\begin{align}
\label{eq:depoCat}
\notag \rho^\mathrm{cat}_\mathrm{eq} \leq \, 1 - &(1 - \rho)(1 - P_\mathrm{syn}) \\ 
    & \times \Big[(1 - \rho_{2Q})(1 - \rho_\mathrm{init}^\mathrm{cat})\Big]^{r \, \delta_\mathrm{max}}
\end{align}
where $\rho$ is the depolarizing error probability on each qubit before the gadget, and $P_\mathrm{syn}$ is the faulty syndrome error probability due to the previous error correction cycle.
Finally, an upper bound on the logical error rate after the gadget can be obtained by equation~\eqref{eq:error_probWithBetaUB} with $\rho$ replaced with $\rho^\mathrm{cat}_\mathrm{eq}$ from equation~\eqref{eq:depoCat}, in conjunction with the results from Section~\ref{sec:MacWilliams_intro}.

\subsection{Flag Error Detection}

In this scenario, a single ancillary qubit is assigned to each syndrome bit. However, to mitigate hook errors, additional flag qubits can be introduced \cite{Cha18:FlagErrCorr, Cha18:FlagsDet}. 
These flag qubits are designed to propagate any hook errors, and their subsequent measurements allow for the detection or the correction of such errors.
Upon detection, the computation can be post-selected and restarted, as an example, to maintain the integrity of the syndrome extraction process.
Error detection and correction flag circuits used for measuring syndrome bits in codes of distance three and five are presented in \cite{Cha18:FlagsDet, Pra23:OtimizedSchemeFlag5and7}.
Here, the probability of syndrome error per single measurement can be upper bounded as 
\begin{align}
\label{eq:depoFlagMeas}
    \notag p_{i} \leq 1 \,- \, &(1 - \rho_\mathrm{init})(1 - \rho_{1Q})^2 \\
    &\times (1 - \rho_{2Q})^{\gamma_i + 2 f_i}(1 - \rho_{\mathrm{meas}})
\end{align}
where $f_i$ denotes the number of flag qubits required for the $i-th$ generator, and $\rho_{\mathrm{init}}$ represents the depolarizing error probability associated with qubit initialization.
Note that each flag qubit is associated with two two-qubit gates involving the syndrome qubit. 
Also, we can compute $P_\mathrm{syn}$ using \eqref{eq:depoFlagMeas} in \eqref{eq:depoCatMeas} and \eqref{eq:MEasERRCAT}.
In this case, the equivalent depolarizing error probability per qubit at the end of the gadget $\rho^\mathrm{flag}_\mathrm{eq}$ is
\begin{align}
\label{eq:depoFlag}
    \rho^\mathrm{flag}_\mathrm{eq} \leq 1 - (1 - \rho)(1 -  P_\mathrm{syn})(1 - \rho_{2Q})^{r \, \delta_\mathrm{max}}
\end{align}
where $\rho$ is the depolarizing error probability on each qubit before the gadget, and $P_\mathrm{syn}$ is the faulty syndrome error probability due to the previous error correction cycle.

Moreover, we must account for the possibility that a flag gadget might fail to detect a hook error, for instance, due to an additional error occurring during its measurement.
Since we are computing an upper bound, we assume that such a failure results in a logical error.
Specifically, the error probability in a flag gadget can be upper bounded as
\begin{align}
\label{eq:depoFlagMeas2}
    \notag
    p_\mathrm{flag} \leq 1 \, - \, &(1 - \rho_\mathrm{init})(1 - \rho_{1Q})^2\\
    &\times (1 - \rho_{2Q})^2(1 - \rho_{\mathrm{meas}})
\end{align}
Given a generator $i$, with $1 \leq i \leq n - k$, and a specific two-qubit gate $c$, with $1 \leq c \leq \gamma_i$, used for measuring that generator, we define $F_{i,c}$ as the number of flag qubits that, in the absence of errors, are capable of detecting a hook error resulting from that particular two-qubit gate. 
This corresponds to all the flag qubits that have been entangled with the syndrome qubit before the two-qubit gate $c$, but have not yet been disentangled at the time $c$ is applied.
Then, the probability that an hook error is undetected is upper bounded by
\begin{align}
\label{eq:depoFlagMeas3}
    P_\mathrm{un,hook} \leq \rho_{2Q} \sum_{i = 1}^{n - k}\sum_{c = 2}^{\gamma_i - 2} p_\mathrm{flag}^{F_{i,c}}.
\end{align}
Note that the inner summation excludes the first and last two two-qubit gates of the generator, as these cannot give rise to hook errors (up to a stabilizer generator).
Finally, an upper bound on the logical error rate can be computed as
\begin{align}
\label{eq:depoFlagLogErr}
    \rho^\prime_L \leq  \rho_L +  P_\mathrm{un,hook} - \rho_L P_\mathrm{un,hook}
\end{align}
where $\rho_L$ is given by equation~\eqref{eq:error_probWithBetaUB} with $\rho$ replaced with $\rho^\mathrm{flag}_\mathrm{eq}$ from equation~\eqref{eq:depoFlag}, in conjunction with the results from Section~\ref{sec:MacWilliams_intro}.

\subsection{Steane Error Correction}
This gadget, specifically designed for CSS codes, involves preparing logical $\ket{0}$ and $\ket{+}$ ancilla states, followed by the transversal application of CNOT gates to propagate $X$ and $Z$ errors from the data block to the ancilla \cite{Ste97:SteaneGadget}. 
The ancilla states are then destructively measured to extract the syndrome. 
Due to the transversality of the CNOT operations, no hook errors can occur, and the circuit does not require repeated measurements.
Here, the ancilla error probability per qubit can be upper bounded as
\begin{align}
\label{eq:STEFlagMeas}
    P_a \leq 1 - (1 - \rho_a)(1 - \rho_{1Q})(1 - \rho_{2Q})(1 - \rho_{\mathrm{meas}})
\end{align}
where $\rho_a$ is the depolarizing error probability on each ancilla qubit before the gadget.
Note that, employing this error correction gadget, a single qubit error in the ancilla state can only produce a single qubit error in the encoded state.  
The depolarizing error probability on each qubit after the gadget is upper bounded as
\begin{align}
\label{eq:STEFlag}
    \rho^\mathrm{Ste}_\mathrm{eq} \leq 1 - (1 - \rho)(1 - \rho_{2Q})^2(1 - P_a)^2
\end{align}
where $\rho$ is the depolarizing error probability on each qubit before the gadget. 
Finally, an upper bound on the logical error rate after the gadget can be obtained by equation~\eqref{eq:error_probWithBetaUB} with $\rho$ replaced with $\rho^\mathrm{Ste}_\mathrm{eq}$ from equation~\eqref{eq:STEFlag}, in conjunction with the results from Section~\ref{sec:MacWilliams_intro}.

\section{Numerical Results}\label{sec:NumRes}

In this section, we evaluate the performance of some codes, in terms of logical error rate $\rho_L$ vs. physical error rate $\rho$, based on the analysis described previously. In particular, the initial four analyses concentrate on assessing the asymptotic logical error rate of certain quantum codes introduced in preceding sections, namely when $\rho \ll 1$, across symmetric and asymmetric channels. Moreover, the concluding analysis addresses the scenario of elevated physical error rates.

\emph{1) Comparison between the $\hat{\beta}_{t+1}$ from the \ac{WE} and the exact $\beta_{t+1}$.}
In Tab.~\ref{tab:paramBeta} we compare the estimated $\hat{\beta}_{t+1}$ from Section~\ref{sec:MacWilliams_B2A}  (equations  \eqref{mac_beta_est_mwpm}, \eqref{mac_beta_est3}, \eqref{mac_beta_est_mwpm_es}), and the exact $\beta_{t+1}$ from Sections~\ref{sec:MacWilliams_B2} and~\ref{sec:exSearch} over the depolarizing channel. 
We remark that the estimates $\hat{\beta}_{t+1}$ are derived solely by employing the logical weight enumerator of the particular quantum code (i.e., from the MacWilliams identities, or \eqref{eq:Ldsurf} for surface codes), whereas the exact values are obtained through a counting argument applied to the logical operators. 
Specifically, both the $[[5,1,3]]$ perfect code and the $[[7,1,3]]$ Steane code exhibit asymptotic non-degeneracy, leading to a congruence between the estimated values and their exact counterparts. 
Instead, the $[[9,1,3]]$ Shor code displays strong degeneracy, so that the estimate is useless ($\hat{\beta}_{2}=0$). 
In this case, we can resort to the logical operator analysis detailed in Section~\ref{sec:MacWilliams_B2} which gives the exact $\beta_2$.
Regarding surface codes, we observe that the estimated $\hat{\beta}_{t+1}$ closely approximate the exact values, with the disparity diminishing as the code's distance is increased.

\input{Figures/cuteTableBeta}

\begin{figure}[tb]
	\centering
	 \resizebox{\columnwidth}{!}{	    
    \includegraphics[width=\columnwidth]{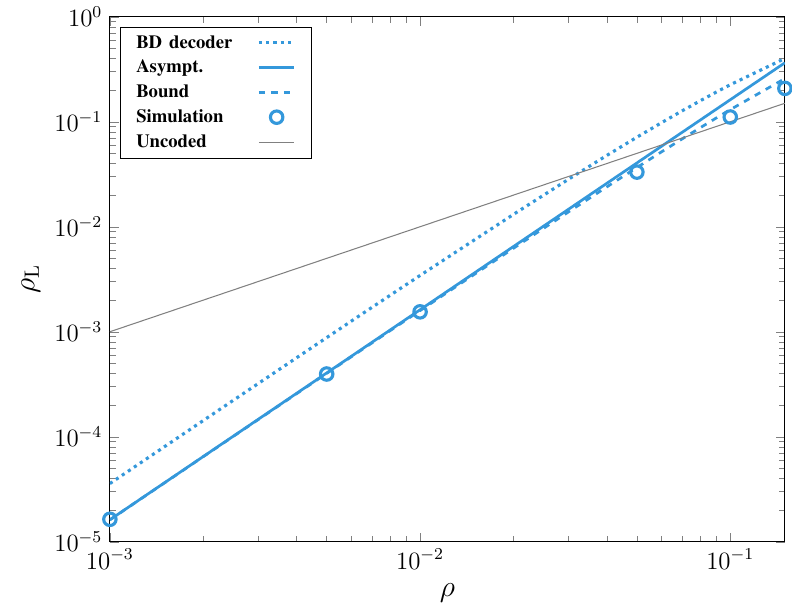}
    }
	\caption{Logical error rate, $[[9,1,3]]$ Shor code over a depolarizing channel. Comparison between theoretical analysis (curves) and simulation (symbols). The curves refer to: the \ac{BD} decoding performance \eqref{eq:PeGen}; the \ac{MW} decoding upper bound \eqref{eq:error_probWithBetaUB} and its asymptotic approximation \eqref{eq:error_probWithBetaApprox} with the exact $\beta_2$ from Tab.~\ref{tab:paramBeta}. 
		\label{Fig:plot_preformanceB2_shor}}
\end{figure}

\emph{2) Comparison between analysis and simulation for the $[[9,1,3]]$ Shor code.} To verify the correctness of the proposed analytical approach we start by studying the $[[9,1,3]]$ Shor code. As observed, for this code the estimation \eqref{mac_beta_est_mwpm} is not useful, as it gives $\hat{\beta}_2 = 0$, which  coincides with the bounded distance decoding due to the strong degeneracy of the Shor code. Therefore, we use the logical operator analysis of Section \ref{sec:MacWilliams_B2}. Fig.~\ref{Fig:plot_preformanceB2_shor} shows, for the [[9,1,3]] code, a comparison between the upper bound \eqref{eq:error_probWithBetaUB}, the asymptotic approximation \eqref{eq:error_probWithBetaApprox} (which becomes \eqref{eq:rhoLShor}), and the logical error rate obtained via Monte Carlo simulations, adopting a \ac{MW} decoder.  
It can be seen that the results are in perfect agreement for $\rho < 0.1$, while there is a small gap for larger $\rho$. This gap arises because \eqref{eq:error_probWithBetaUB} implicitly assumes $\beta_j=0$ for $j\geq 3$, while the Shor code is able to correct also a little percentage of errors of weight $j \geq 3$, as for instance  $\M{X}$ errors in three different qubit triplets ($\M{X}_1\M{X}_4\M{X}_7$). Moreover, in the plot, we report the  error probability with the \ac{BD} decoder, computed using \eqref{eq:PeGen}, which has the same trend as the \ac{MW} decoder. The gap between the two curves is due to the fraction of weight two errors which are corrected by the \ac{MW} decoder.  

\begin{figure}[tb]
	\centering
	\resizebox{\columnwidth}{!}{
        \includegraphics{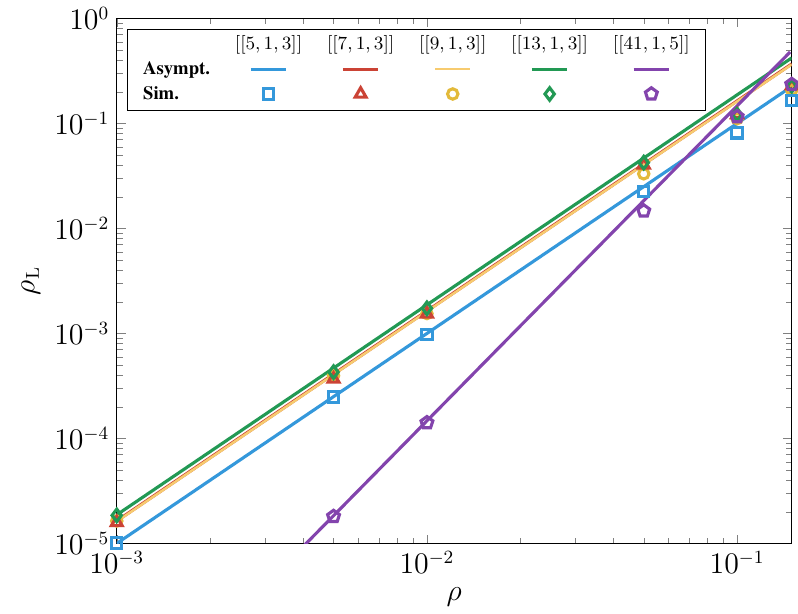}
	}
	\caption{Logical error rate vs.\ physical error rate, $[[5,1,3]]$ code, $[[7,1,3]]$ Steane code, $[[9,1,3]]$ Shor code, the  $[[13,1,3]]$ surface code, and the $[[41,1,5]]$ surface code, over a depolarizing channel. Comparison between theoretical analysis (curves) and simulation (symbols). The solid curves refer to the asymptotic approximation \eqref{eq:error_probWithBetaApprox} with the exact $\beta_{t+1}$ from Tab.~\ref{tab:paramBeta}.}
		\label{Fig:plot_preformanceB2}
\end{figure}

\emph{3) Asymptotic performance analysis over depolarizing channel.} 
In Fig.~\ref{Fig:plot_preformanceB2}, we present a comparison between the asymptotic approximation of the logical error rate and the results obtained from Monte Carlo simulations using a \ac{MW} decoder.
The analysis includes the $[[5,1,3]]$ perfect code, the $[[7,1,3]]$ Steane code, the $[[9,1,3]]$ Shor code, the $[[13,1,3]]$ surface code, and the $[[41,1,5]]$ surface code.
In doing so, we use \eqref{eq:error_probWithBetaApprox} and the values of $\beta_2$ obtained in Section~\ref{sec:MacWilliams_B2}. 
In particular, we can see that the $[[5,1,3]]$ perfect code has the best error correction capability among the codes with distance $d = 3$, despite  it is not able to correct any error of weight $j \geq 2$. This is because it is shorter than the others, so less prone to channel errors. For the same reason, Steane and Shor codes show almost the same performance even if the former has a much smaller value of $\beta_2$. Finally, surface codes pay the price of all the implementation benefits that their lattice provides. For instance, even if the $[[13,1,3]]$ surface code is able to correct many  errors of weight $j = 2$, it uses a large number of physical qubits, resulting in worst performance than the previous codes. 

\begin{figure}[tb]
	\centering
	\resizebox{\columnwidth}{!}{
     \includegraphics{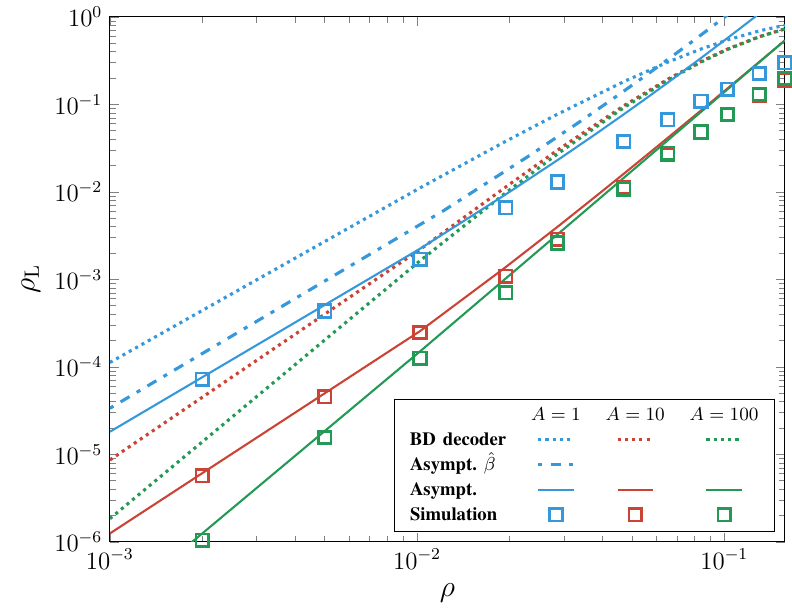}
	}
	\caption{Logical error rate, $[[23,1,3/5]]$ surface code over symmetric and asymmetric channels. Comparison between: the \ac{BD} decoding performance \eqref{eq:PeAsym}; the asymptotic approximation \eqref{eq:betaj_asym} with the estimated $\hat{\beta}_2$ and $\hat{\beta}_3$ from Tab.~\ref{tab:paramBeta}, and with their exact values from Tab.~\ref{tab:Err}; the simulations with a \ac{MWPM} decoder.}
		\label{Fig:plot_asymm_3x5}
\end{figure}

\emph{4) Asymptotic performance analysis over asymmetric channels.} In Fig.~\ref{Fig:plot_asymm_3x5}, we study the performance over channels with asymmetry parameter $A=1$,  $A = 10$, and $ A=100$, for the $[[23,1,3/5]]$ surface code.  Specifically, we have determined the bounded distance performance using \eqref{eq:PeAsym}. 
It is interesting to note that, for all kinds of bias of the channel, the simulated logical error rate has the same behavior of the bound error probability, but with a gap between each couple of curves. This gap is due to the capability of surface codes to correct many errors of weight $w \geq t + 1$. However, since not all the errors of weight $w = t + 1$ can be corrected, we have $\beta_{t+1}>0$, and this 
makes the asymptotic slope to be $t+1$, no matter how small is $\beta_{t+1}$. Moreover, we computed the asymptotic approximations of the logical error rate for the $[[23,1,3/5]]$ surface code, over the same symmetric and asymmetric channels using \eqref{eq:betaj_asym}. The values of $\beta_j$ used in this figure are shown in Tab.~\ref{tab:Err}. We can observe that, for $\rho \ll 1$, these curves are tight to the respective Monte Carlo simulations.
Finally, in the case of a depolarizing channel, we provide the approximation obtained with the estimated $\hat{\beta}_j$ from MacWilliams identities, computed by \eqref{mac_beta_est_mwpm23A} and \eqref{mac_beta_est_mwpm23B}. Notably, this closely approximates the actual performance of the code.

\begin{figure}[t]    
 \centering 
 \resizebox{\columnwidth}{!}{
 \includegraphics[width=\columnwidth]{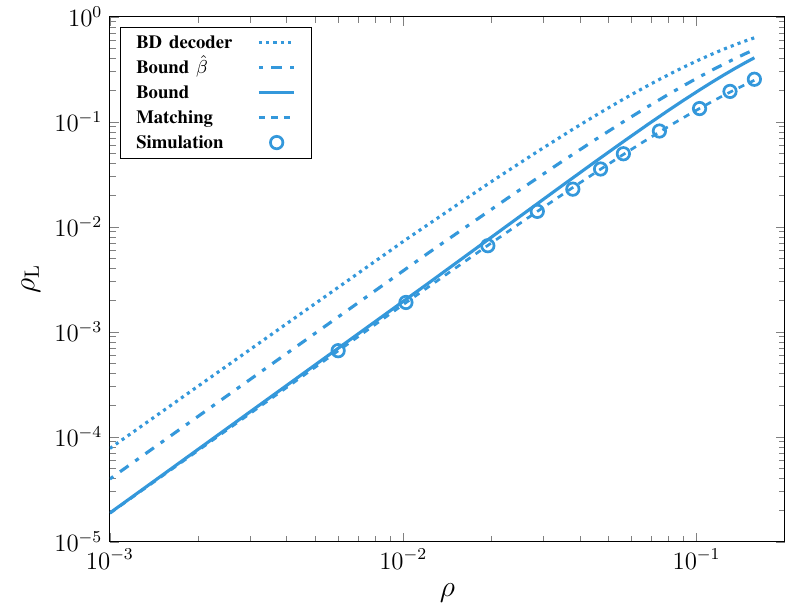}
 }
\caption{Logical error rates, comparison between theoretical analysis and simulation, \ac{MWPM} decoder:   $[[13,1,3]]$ surface code over a depolarizing channel. The curves refer to: the \ac{BD} decoding performance \eqref{eq:PeGen}; the upper bound \eqref{eq:error_probWithBetaUB} with the estimated $\beta_2$ and its exact value from Tab.~\ref{tab:paramBeta}; the matching \eqref{eq:rho_L} using $(\beta_2, \beta_3, \beta_4, \beta_5, \beta_6) = (0.76, 0.48, 0.48, 0.46, 0.5)$.}
\label{Fig:plot_3x3_matching}
\end{figure}

\begin{figure}[t]    
 \centering 
 \resizebox{\columnwidth}{!}{
    \includegraphics[width=\columnwidth]{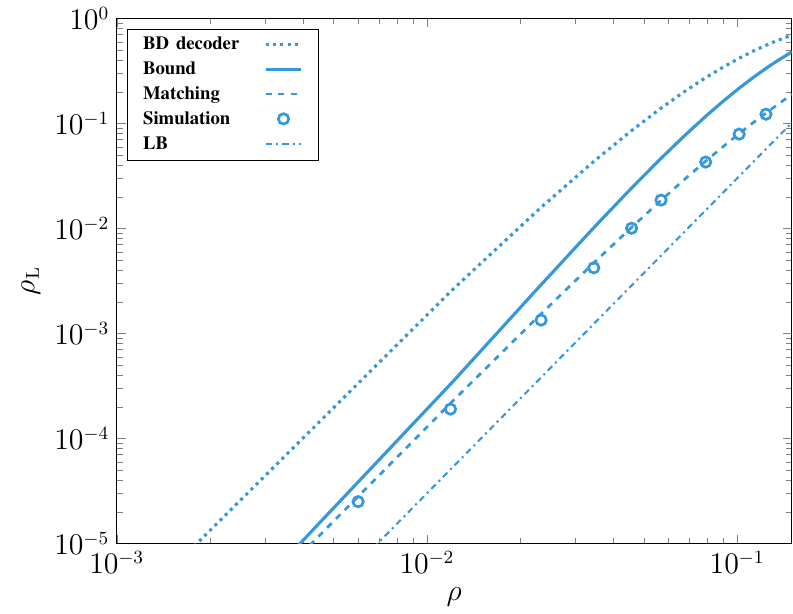}
 }
\caption{Logical error rates, comparison between theoretical analysis and simulation, \ac{MWPM} decoder: $[[23,1,3/5]]$ surface code over a phase flip channel. The curves refer to: the \ac{BD} decoding performance \eqref{eq:PeGen}; the upper bound \eqref{eq:error_probWithBetaUB} with the exact value from Tab.~\ref{tab:Err}; the matching \eqref{eq:rho_L} using $\left(\beta_3, \beta_4, \beta_5, \beta_6, \beta_7\right) = (0.92, 0.76, 0.59, 0.52, 0.49)$. The dash-dotted curve re-presents the asymptotic lower bound from \cite{Lee:21}.}
\label{Fig:plot_3x5_matching}
\end{figure}

\emph{5) High error-rate performance analysis.}
We want here to show that, by using \eqref{eq:rho_L} with several $\beta_j$ besides $\beta_{t+1}$, it is possible to estimate the code performance not only for small $\rho$ (for this the $\beta_{t+1}$ would suffices), but also for larger values of $\rho$. In Fig.~\ref{Fig:plot_3x3_matching} and Fig.~\ref{Fig:plot_3x5_matching}
we show the performance as given by equation \eqref{eq:rho_L} over depolarizing and phase flip channels, taking into account many values of $\beta_{j}$. To this aim, we compare the analytical formulas with simulations for the $[[13,1,3]]$ surface code over a depolarizing channel, and for the $[[23,1,3/5]]$ surface code over a phase flip channel. In Fig.~\ref{Fig:plot_3x5_matching} we also report the asymptotic lower bound, valid only over phase flip channel, from \cite{Lee:21}. 

Specifically, for the $[[13,1,3]]$ surface code over depolarizing channel, we plot \eqref{eq:rho_L} using $(\beta_2, \beta_3, \beta_4, \beta_5, \beta_6) = (0.76, 0.48, 0.48, 0.46, 0.5)$ where $(\beta_2, \beta_3)$ are computed by exhaustive search according to \eqref{eq:beta_j}, while $(\beta_4, \beta_5, \beta_6)$ are approximated using the values computed for $A \to \infty\,$. The $\beta_j$ with $j>6$ are set to zero.
Similarly, for the $[[23,1,3/5]]$ surface code over a phase flip channel,  we used $\left(\beta_3, \beta_4, \beta_5, \beta_6, \beta_7\right) = (0.92, 0.76, 0.59, 0.52, 0.49)$, and $\beta_j=0$ for $j>7$. In both instances, we have seen that  with five coefficients the resulting curves exhibited already a close correspondence to the actual logical error rate across all values of physical error probability, while computing the others $\beta_j$ would be computationally expensive without appreciable improvements in the approximation.
In the figures, we also plot their upper bound \eqref{eq:error_probWithBetaUB}, with $\beta_2=0.76$ and $\beta_3=0.92$ for the $[[13,1,3]]$ and the $[[23,1,3/5]]$ codes, respectively. 
Furthermore, in the case of the $[[13,1,3]]$ surface code, we furnish both the computed bounded distance performance using \eqref{eq:PeGen} and the upper bound on the logical error rate, given by \eqref{eq:error_probWithBetaUB} with \eqref{mac_beta_est_mwpm13}. It's noteworthy that the straightforward upper bound obtained solely through the \ac{WE} exhibits a close alignment with the exact performance. 
As expected, our approximations are tight for $\rho \ll 1$, allowing to estimate logical error rates not achievable by Monte Carlo simulations.

\begin{figure}[t]    
 \centering 
 \resizebox{\columnwidth}{!}{
\includegraphics[width=\columnwidth]{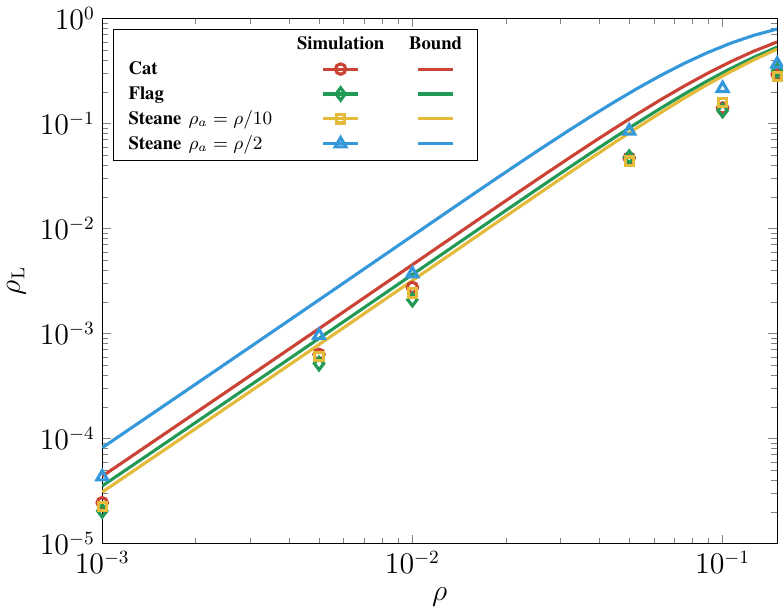}
 }
\caption{Logical error rates considering noisy syndrome extraction: comparison between theoretical analysis and simulation using the \ac{MWPM} decoder for the $[[13,1,3]]$ surface code over a depolarizing channel, assuming uniform noise parameters $\rho_{2Q} = \rho_{1Q} = \rho_\mathrm{init} = \rho_\mathrm{init}^\mathrm{cat} = \rho_{\mathrm{meas}} = \rho / 100$. }
\label{Fig:noisySyndrome}
\end{figure}
\begin{figure}[t]    
 \centering 
 \resizebox{\columnwidth}{!}{
\includegraphics[width=\columnwidth]{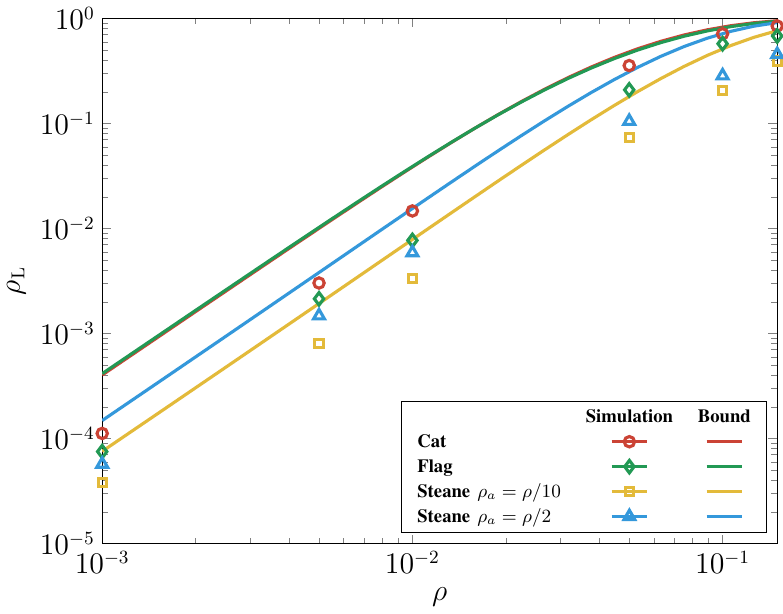}
 }
\caption{Logical error rates considering noisy syndrome extraction: comparison between theoretical analysis and simulation using the \ac{MWPM} decoder for the $[[13,1,3]]$ surface code over a depolarizing channel, assuming uniform noise parameters $\rho_{2Q} = \rho_{1Q} = \rho_\mathrm{init} = \rho_\mathrm{init}^\mathrm{cat} = \rho_{\mathrm{meas}} = \rho / 10$.}
\label{Fig:noisySyndrome_10}
\end{figure}
\begin{figure}[t]    
 \centering 
 \resizebox{\columnwidth}{!}{
\includegraphics[width=\columnwidth]{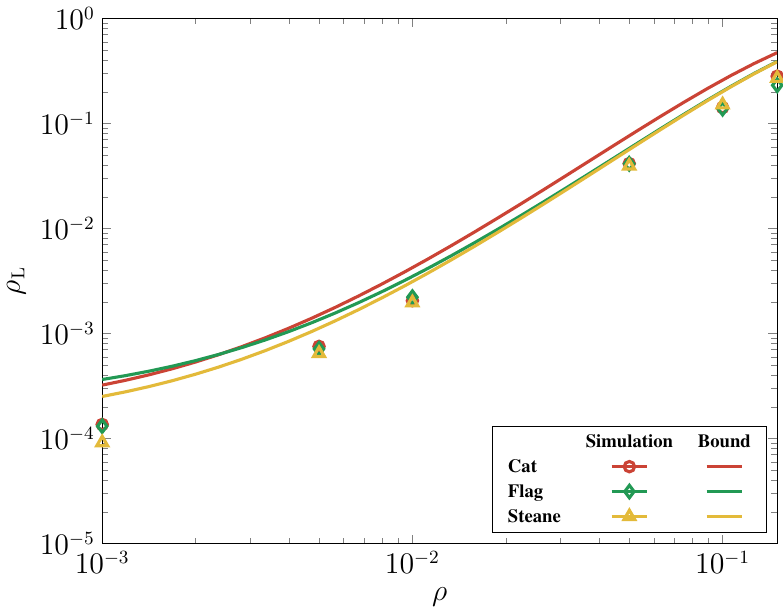}
 }
\caption{Logical error rates considering noisy syndrome extraction: comparison between theoretical analysis and simulation using the \ac{MWPM} decoder for the $[[13,1,3]]$ surface code over a depolarizing channel, assuming noise parameters $\rho_{2Q} = \rho_\mathrm{init} = \rho_\mathrm{init}^\mathrm{cat} = \rho_{\mathrm{meas}} = 10^{-4}$, $\rho_{1Q} = 2\cdot 10^{-5}$, and $\rho_a = 10 \rho_{2Q}$.}
\label{Fig:noisySyndrome_fixed}
\end{figure}

\emph{6) Noisy syndrome extraction performance analysis.} In Fig.~\ref{Fig:noisySyndrome} and Fig.~\ref{Fig:noisySyndrome_10}, for the $[[13,1,3]]$ surface code, we compare the simulated logical error rates under noisy syndrome extraction with the upper bounds described in Section~\ref{Sec_syndromeExtr}.
In Fig.~\ref{Fig:noisySyndrome} we set the error rates as $\rho_{2Q} = \rho_{1Q} = \rho_\mathrm{init} = \rho_\mathrm{init}^\mathrm{cat} = \rho_{\mathrm{meas}} = \rho / 100$, while in Fig.~\ref{Fig:noisySyndrome_10} we set $\rho_{2Q} = \rho_{1Q} = \rho_\mathrm{init} = \rho_\mathrm{init}^\mathrm{cat} = \rho_{\mathrm{meas}} = \rho / 10$. 
Furthermore, we set the Steane ancillary state preparation error $\rho_a = \rho / 10$, or $\rho_a = \rho / 2$, to ensure that the ancillary resource state employed for syndrome extraction is more reliable than the data state it is used to correct.
To achieve fault-tolerant syndrome extraction, one strategy is to repeatedly measure all stabilizer generators until the same syndrome is obtained $t + 1$ times consecutively. 
Considering that up to $t$ errors may occur during this process, in the worst-case scenario, $(t + 1)^2$ measurement rounds are required~\cite{Sho96:CatStates, Got:96}.
Hence, we choose $r = (t + 1)^2 = 4$, while $\delta_{\max} = 4$, since some qubits participate in two $\M{X}$ and two $\M{Z}$ generators.
For weight-four generators, two flags ancillas are needed for fault detection.
Setting $\beta_2 = 0.76$, we compute upper bounds on the logical error rate by applying equations~\eqref{eq:depoCat}, \eqref{eq:depoFlag}, and \eqref{eq:STEFlag} in~\eqref{eq:error_probWithBetaUB}, for the various gadgets discussed in Section~\ref{Sec_syndromeExtr}.  
From our analysis, we observe that the proposed upper bounds are tight with respect to the numerical simulations.
For circuit error rates equal to $\rho / 100$ and $\rho_a = \rho / 10$, the three techniques exhibit comparable performance.
In particular, the effectiveness of the Steane syndrome extraction gadget is highly dependent on the fidelity of the resource state used in the procedure.
Specifically, when $\rho_a = \rho / 10$, it achieves the lowest logical error rates among the considered gadgets; however, for $\rho_a = \rho / 2$, its performance deteriorates significantly.
In contrast, the performance of the cat and flag syndrome extraction gadgets depends strongly on the circuit error rates: when these are set to $\rho / 10$, the resulting logical error rates are substantially higher.

In Fig.~\ref{Fig:noisySyndrome_fixed}, we present a comparison of simulated logical error rates for the $[[13,1,3]]$ surface code under noisy syndrome extraction with realistic noise parameters, alongside the upper bounds introduced in Section~\ref{Sec_syndromeExtr}.
Based on recent progress in trapped-ion technology, which exhibits some of the lowest physical error rates, we set the relevant error parameters to $\rho_{2Q} = \rho_\mathrm{init} = \rho_\mathrm{init}^\mathrm{cat} = \rho_{\mathrm{meas}} = 1\cdot 10^{-4}$, and $\rho_{1Q} = 2\cdot 10^{-5}$ \cite{ Brown11_SingleQubit, Erickson22:measurement, hug25:2qubit}.
Additionally, we take $\rho_a = 10\rho_{2Q}$ to account for the gates required in the preparation of the ancillary resource state used in Steane syndrome extraction \cite{ForAma25:FTCSS}.
We observe that quantum error correction remains effective only as long as gate error rates are significantly lower than the errors it is designed to protect against.

\section{Conclusions}\label{sec:conclusions}

The aim of this work was to propose theoretical bounds on the error correction capability of \ac{CSS} stabilizer quantum codes, providing guidance for the development of future fault-tolerant quantum systems once the physical qubit error rate and required reliability are established. 
These bounds were obtained starting from the \acl{WE} for undetectable errors, which was derived starting from the quantum MacWilliams identities.
Furthermore, we proposed a method to achieve even more stringent bounds by analyzing the structure of the logical operators for a specific stabilizer code. 
As an example of application of the method, we examined the performance of some stabilizer codes with \ac{MW} decoding, as well as surface codes with \ac{MWPM} decoding, comparing the theoretical results with simulations on both symmetric and asymmetric quantum channels. 
Notably, we have made advances in understanding quantum degeneracy.
Indeed, we have shown that the asymptotic degeneracy of a quantum code is strongly related to the fraction of errors of weight $t+1$ that are shared between more than one logical operator of weight $2t + 1$ and $2t + 2$.
For instance, we can observe that the Steane code, while being degenerate, can be regarded asymptotically non degenerate.
This is because there are no logical operators of weight $2t+1$ overlapping with Pauli patterns of weight $t+1$.
Conversely, the Shor code is strongly asymptotically degenerate, as a number of $\M{Z}\M{Z}\M{Z}$ logical operators share a pair of Pauli $\M{Z}$.
This is closely tied to the performance of these codes.
Indeed, the Shor code, being able to correct a larger number of errors beyond the code's distance, exhibits superior error correction capability.
This fact provides a useful criterion in the design of quantum codes where logical operators of minimum weight share error patterns greater than the guaranteed error correction capability of the code. 
Finally, by extending the analysis to include realistic noisy syndrome extraction circuits, we provide a method for bounding logical error rates under circuit-level noise models. 
This approach to the theoretical analysis of noisy circuits paves the way for an in-depth evaluation of fault-tolerant quantum error correction performance.


\section*{Acknowledgment}

This work was partially supported by the European Union under the Italian National Recovery and Resilience Plan (NRRP) of NextGenerationEU, HPC National Centre for HPC, Big Data and Quantum Computing (CN00000013).


\section*{Appendix}\label{app:boundCSSDC}
Here we provide the detailed derivation of \eqref{mac_beta_est3}. Specifically, for a general \ac{CSS} code, a logical operator made of $\M{Y}$ Pauli operators can be caused only by channel errors of weight $t+1$ composed exclusively by $\M{Y}$ operators. This is a consequence to the fact that half of the generators consists of $\M{X}$ Pauli operators while the other half of $\M{Z}$ operators. For instance, considering a \ac{CSS} code of distance $d = 3$, we have that one channel error of weight $t +1 = 2$ that is not exclusively composed by $\M{Y}$ Pauli operators, such as $\M{Z}\M{Y}$, will never cause a $\M{Y}\M{Y}\M{Y}$ logical operator. In fact, the $\M{X}$ error is always corrected by a code of distance $d = 3$, while the remaining $\M{Z}\M{Z}$ error will cause a logical operator composed by $\M{Z}\M{Z}\M{Z}$. Furthermore, assuming no knowledge about the fraction of $\M{Y}$ logical operators, we conclude that a generic logical operator of weight $w=2t + 1$ or  $w=2t + 2$ can be caused by $2^{t+1}$ different Pauli errors of weight $t+1$. Moreover, a \ac{CSS} dual-containing quantum code is characterized by a set of generators made of all $\M{X}$ or $\M{Z}$ Pauli operators applied to the same qubits. As a result, logical operators of minimum weight are made of only one kind of Pauli, i.e., there is an equal number of logical operator composed only of $\M{X}$, of  $\M{Z}$, or of $\M{Y}$. Therefore, for the \ac{CSS} dual-containing, two third of logical operators made either of $\M{X}$ or $\M{Z}$ Pauli matrices are caused by $2^{t+1} - 1$ different Pauli errors, where the $-1$ is to exclude the error composed by $t+1$ $\M{Y}$ operators. Note that this kind of error is the only one that can cause a logical operator composed of only $\M{Y}$ Pauli operators. 

\bibliographystyle{quantum}
\bibliography{Files/IEEEabrv,Files/StringDefinitions,Files/StringDefinitions2,Files/refs}

\end{document}

%% file: Files/Acronimi_SICMMA.tex
\begin{acronym}
\small
\acro{AWGN}{additive white Gaussian noise}
\acro{BCH}{Bose–Chaudhuri–Hocquenghem}
\acro{CDF}{cumulative distribution function}
\acro{CRC}{cyclic redundancy code}
\acro{LDPC}{low-density parity-check}
\acro{ML}{maximum likelihood}
\acro{MWPM}{minimum weight perfect matching}
\acro{QECC}{quantum error correcting code}
\acro{QLDPC}{quantum low-density parity-check}
\acro{PDF}{probability density function}
\acro{PMF}{probability mass function}
\acro{MPS}{matrix product state}
\acro{WEP}{weight enumerator polynomial}
\acro{WE}{weight enumerator}
\acro{BD}{bounded distance}
\acro{CSS}{Calderbank Shor Steane}
\acro{MW}{minimum weight}

\end{acronym}

%% file: Figures/cute_table3bis.tex
\renewcommand{\arraystretch}{1.2}
\begin{table}[t]
    \centering
    \setlength{\tabcolsep}{2pt}
    \caption{Coefficients for performance evaluation, $[[13,1,3]]$ surface code.}
    \label{tab:param}
    \small
    \includegraphics[width=\columnwidth]{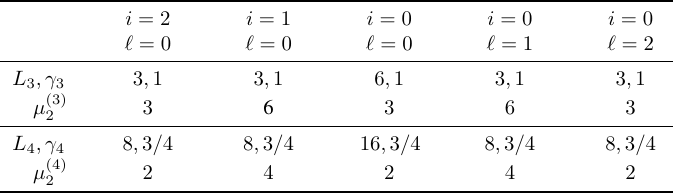}
\end{table}

%% file: Figures/cuteTable.tex
\begin{table*}[t]
    \centering
    \setlength{\tabcolsep}{1pt}
    \caption{Surface codes: fraction of non-correctable error patterns per error class using MWPM.}
    \label{tab:Err}
    \small

\includegraphics[width = \textwidth]{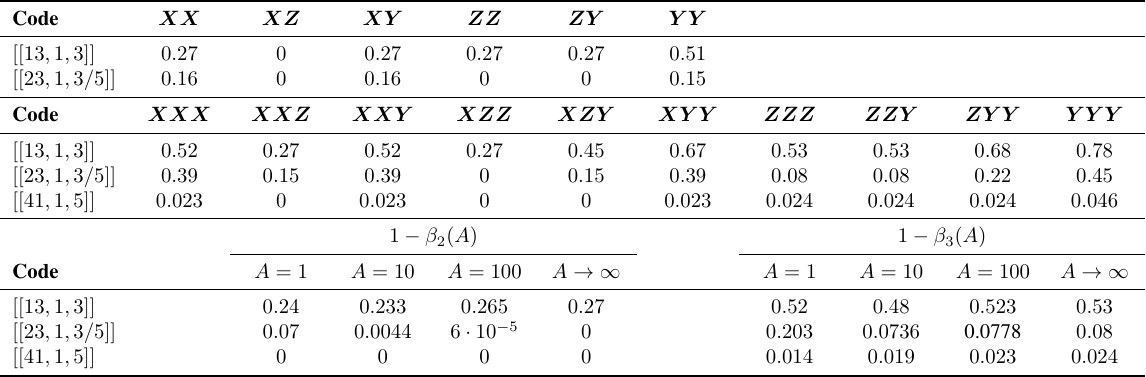}
    
\end{table*}

%% file: Figures/cuteTableBeta.tex
\renewcommand{\arraystretch}{1.2}
\begin{table}[t]
    \centering
    \setlength{\tabcolsep}{2pt}
    \caption{Comparison between the bounds from Section~\ref{sec:MacWilliams_B2A}, and the exact values from Sections~\ref{sec:MacWilliams_B2} and~\ref{sec:exSearch}, depolarizing channel.}
    \label{tab:paramBeta}
    \includegraphics[width = \columnwidth]{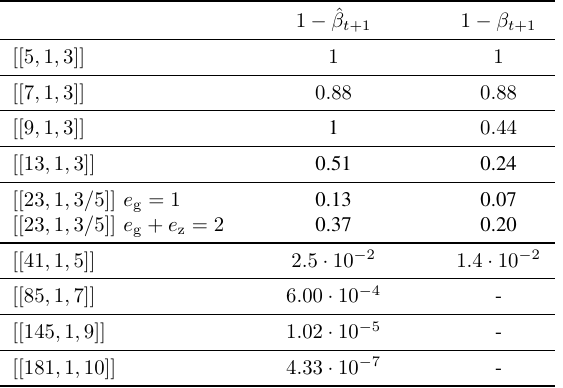}
\end{table}